\documentclass[journal,12pt,onecolumn,draftclsnofoot,]{IEEEtran}

\IEEEoverridecommandlockouts
\usepackage{times,enumerate} 
\usepackage{color}
\usepackage{graphicx}
\usepackage{setspace}
\usepackage{bbm}
\usepackage{mathdots,mathrsfs}
\usepackage{amssymb,latexsym,amsfonts,amsmath,cite,comment,relsize}
\usepackage{stmaryrd}
\usepackage[dvips]{psfrag}
\usepackage{setspace}
\usepackage{amsmath}
\usepackage{url}
\usepackage{soul}
\usepackage{pgf,tikz}
\usepackage[linesnumbered,ruled]{algorithm2e} 
\usepackage{cite}
\usepackage{amsmath,amssymb,amsfonts}
\usepackage{algorithmic}
\usepackage{graphicx}
\usepackage{textcomp}
\usepackage{lineno,hyperref} \usepackage{graphics} 
\usepackage{epsfig} 
\usepackage{mathptmx} 
\usepackage{subfigure}
\usepackage{float}
\usepackage{graphics} 
\usepackage{booktabs}
\usepackage{stackengine}
\usepackage{times} 
\usepackage{amsmath} 
\usepackage[dvips]{psfrag}
\usepackage{psfrag}
\usepackage{optidef}
\usepackage{url} 
\usepackage{mathtools}
\usepackage{tikz}
\usetikzlibrary{shapes,arrows}
\usepackage{verbatim}
\usepackage{standalone} 
\usetikzlibrary{positioning}
\usepackage{optidef}
\usepackage{mathtools, nccmath}

\newtheorem{lemma}{Lemma}
\newtheorem{corollary}{Corollary}

\newtheorem{assumption}{Assumption}

\newtheorem{pf}{Proof}

\DeclareMathOperator{\EE}{\mathcal{E}}
\DeclareMathOperator{\VV}{\mathcal{V}}

\DeclareMathOperator{\T}{\rm T}
\DeclareMathOperator{\diag}{\mathrm{diag}}

\DeclareMathOperator{\HH}{\mathcal{H}}

\newcommand{\V}{{\VV}}

\DeclareMathOperator{\N}{\mathnormal{n}}

\DeclareMathOperator*{\argmax}{arg\,max}
\newcommand{\rhoo}{\rho_\text{ss}}

\title{ \LARGE \bf
	Centrality-Based Traffic Restriction in Delayed Epidemic Networks 
}

\author{Atefe Darabi$^{1}$ and Milad Siami$^{1}$%
\thanks{$^{1}$Department of Electrical \& Computer Engineering,
Northeastern University, Boston, MA 02115 USA
	(e-mails: {\tt\small \{darabi.a, m.siami\}@northeastern.edu}).}%
}

\begin{document}

\maketitle
\allowdisplaybreaks
\begin{abstract}
During an epidemic, infectious individuals might not be detectable until some time after becoming infected. The studies show that carriers with mild or no symptoms are the main contributors to the transmission of a virus within the population. The average time it takes to develop the symptoms causes a delay in the spread dynamics of the disease. When considering the influence of delay on the disease propagation in epidemic networks, depending on the value of the time-delay and the network topology, the peak of epidemic could be considerably different in time, duration, and intensity. 
Motivated by the recent worldwide outbreak of the COVID-19 virus and the topological extent in which this virus has spread over the course of a few months, this study aims to highlight the effect of time-delay in the progress of such infectious diseases in the meta-population networks rather than individuals or a single population. In this regard, the notions of epidemic network centrality in terms of the underlying interaction graph of the network, structure of the uncertainties, and symptom development duration are investigated to establish a centrality-based analysis of the disease evolution. A convex traffic volume optimization method is then developed to control the outbreak. The control process is done by identifying the sub-populations with the highest centrality and then isolating them while maintaining the same overall traffic volume (motivated by economic considerations) in the meta-population level. The numerical results, along with the theoretical expectations, highlight the impact of time-delay as well as the importance of considering the worst-case scenarios in investigating the most effective methods of epidemic containment.
\end{abstract}

%

\section{Introduction}

The large-scale spread of an infectious disease occurs every few years and leads into serious crises before it eventually dies out \cite{qiu2017pandemic}. The extent through which a high-speed epidemic continues depends mostly on first, the government interventions, and second, the existence of an effective treatment against the disease. In this regard, the study of epidemic propagation by network models has been theoretically proven to be useful in determining the most effective methods of epidemic management and control as well as allocating treatment and immunization resources \cite{xu2009effect, zanette2002effects}.

The macro-modeling (or meta-population) representation of the epidemic networks is studied in \cite{bailey1986}, where the author introduces the general form of the spread dynamics in a community of sub-populations. Many studies have been using meta-population network of Susceptible-Infected-Susceptible (SIS) \cite{preciado2013}, Susceptible-Infected-Removed (SIR) \cite{terry2010, lioyd2004}, and Susceptible-Exposed-Infected-Removed (SEIR) \cite{lioyd2004} models to develop the dynamics of the epidemic in macro scale.

The centrality analysis of a time-delayed consensus network has been studied in \cite{ghaedsharaf2019centrality} and a closed form of the node and link centralities in the presence of different noises in the network is offered. In \cite{Olfati:2004}, the authors have studied the convergence analysis of the time-delayed networks with linear dynamics. It is shown that the convergence rate of delayed networks with linear dynamics has a correlation with time-delay and the largest eigenvalue of the graph. The stochastic delayed SIS model has been exploited in \cite{liu2019} to find the threshold behavior of the systems with vaccination and double diseases.

Significant work has been done in the area of epidemic elimination, disease spread control, and designing public health control measures against the epidemics \cite{zaric2001resource, zaric2002dynamic, ferretti2020quantifying}. Some suggest that an optimized alternation in the disease network structure could hinder the disease effectively. Node or edge removal is the direct way of modifying the network structure which is also called spectral control, as it aims to minimize the largest eigenvalue of the network to locally stabilize it around the disease-free equilibrium \cite{preciado2010, preciado2013}. However, the spectral radius minimization has been proven to be an NP-hard problem \cite{van2011decreasing}. Network metric-based node or edge removal is another common approach in spectral control. A removal strategy based on the number of the closed walks and assortativity effect has been proposed in \cite{van2011decreasing}.

On the other hand, vaccination and traffic flow restriction modify the epidemic recovery and infection rates, which in turn, affect the disease progress pattern. Various control strategies have been proposed in this area. A geometric programming-based approach has been presented in \cite{preciado2013} for the optimal resource allocation in a meta-population network. In this study, the infection and recovery rates are assumed to be the optimization variables. The intercity traffic restriction of an SIS mate-population model has also been studied in \cite{preciado2013traffic}. Optimal and heuristic local solutions are proposed to completely eradicate the disease, where the cost of the second solution is shown to be higher than the cost of the first one. In terms of social awareness effect on the epidemic evolution, \cite{preciado2013convex} has proposed a semidefinite programming-based optimization to control a network that follows the Susceptible-Alert-Infected-Susceptible (SAIS) epidemic model.

In the real world cases, node removal happens by social distancing, quarantining, vaccinating individuals, or complete shut down of certain areas in the city \cite{nowzari2016}. In this regard, \cite{miller2007} has introduced PageRank-based vaccination strategy for realistic social networks. The authors in \cite{holme2002attack} have also proposed a network attack strategy based on degree and betweenness centrality for both passive and online node and edge removal approaches. A Pulse vaccination strategy has been offered in \cite{terry2010} which can eradicate the disease in all the sub-populations.

In 2020, many network scientists have established their study on the modeling and control of COVID-19, as it has turned into a global crisis. In the recent research study \cite{morris2020optimal}, the optimal intervention starting time and duration that will reduce the epidemic peak and length in an SIR epidemic model has been proposed. The effect of a complete suppression, mistimed interventions, and sustained interventions have also been investigated in this study \cite{morris2020optimal}.
The effect of social distancing start time, period, and duration on the infection peak of COVID-19 in epidemic models has also been investigated in \cite{sadeghi2020universal}. It is shown that a well-timed intervention improves flattening the infection curve in various epidemic models, such as SIR, feedback-SIR (fSIR), Susceptible-Infected-Quarantined-Susceptible (SIQR), Susceptible-Infected-Diagnosed-Ailing-Recognized-Threatened-Healed-Extinct (SIDARTHE), and 6 compartment SIR \cite{sadeghi2020universal}. 

A more generalized SEIR model of COVID-19 nationwide spread in China including self-protection and quarantine compartments has been established to predict the disease propagation pattern and to find a possible extinction time for some of the provinces \cite{peng2020epidemic}. An SIRD model has been developed to estimate the COVID-19 epidemic parameters in Italy, using real-data and a non-convex parameter identification approach \cite{calafiore2020modified}. An agent-based model of COVID-19 dynamics is utilized to simulate the multi-wave behavior of the spread while applying different possible government interventions such as testing and tracing, and travel restriction on it \cite{rajabi2020investigating}.

\noindent{\bf Our contributions.} Inspired by the recent COVID-19 outbreak, this study is dedicated to the epidemiological investigation of the infectious diseases with the following main contributions:
\begin{itemize}
      \item[i.] Modeling the time-delayed epidemic dynamics of meta-populations using the network SIS model (Section \ref{sec2}).
    \item[ii.] Investigating the effect of time-delay and different sources of uncertainty on the modeled network dynamics (Sections \ref{sec2} and \ref{sec3}).
    \item[iii.] Developing the explicit centrality measure of sub-populations with respect to time-delay and transportation network structure (Section \ref{sec3}).
    \item[iv.] Designing optimal and robust methods of traffic restriction based on the steady-state performance of the network (Section \ref{sec4}).
\end{itemize}

The simulation results for a core-periphery network and also the network of United States busiest airports are presented in Section \ref{sec5}.

\section{Delayed epidemic network of meta-population models} \label{sec2}
The proportion of pre-symptomatic individuals in the population could play an important rule in the disease propagation pattern. In the case of COVID-19 for instance, studies indicate that a significant proportion of positive tests belongs to the pre-symptomatic individuals who are just the carriers \cite{WHO2020_46:Online} and might not develop any symptoms up to 14 days \cite{WHO2020_73:Online}. The effect of pre-symptomatic individuals in the spread of the disease can be modeled as a time-delay in the network dynamics. In this study, the average time of symptom development for all the individuals is assumed to be identical and equal to a non-negative constant $\tau$.  

\subsection{Deterministic meta-population SIS model with time-delay}
Let the undirected and weighted graph $\mathcal{G = (V, E,} w)$ represent a meta-population in the epidemic network. $\VV=\{1,2,\ldots,n\}$ shows the set of nodes in the graph or the group of cities, states, countries, or in general, sub-populations in the epidemic network. $\EE$ denotes the edge set which shows the connection between every two member of $\VV$ with the corresponding weight of $w_e = a_{ij}$ for all $e= \{i,j\} \in \EE$ for $i\neq j$. The counterpart of weighted edges in the epidemic network would be the transportation capacity or traffic volume between every two sub-population in $\VV$. In addition to its neighbors, the infection state of a sub-population depends on the progress of disease within the sub-population as well, which itself is a function of various factors such as social distancing. Let us project the final effect of social distancing in the sub-population $i$ on parameter $a_{ii}$, where $a_{ii} = 0$ belongs to a sub-population which follows social distancing and as $a_{ii}$ increases, the social distancing rules become less strict. The adjacency matrix of the corresponding graph is then defined as,
\begin{equation} \label{eq:A}
     A = \\
    \begin{bmatrix}
    a_{11} & a_{1 2} & \cdots & a_{1 n}\\
    a_{2 1} & a_{22} & \cdots & a_{2 n}\\
    \vdots & \vdots & \ddots & \vdots\\
    a_{n 1} & a_{n 2} & \cdots & a_{nn}
    \end{bmatrix}. 
\end{equation}

The state of the epidemic network at time $t\geq0$ is represented by the vector $\mathbf{p}(t) =
\left[p_1 (t),\ldots,p_n (t)\right]^{\rm{T}}$, where $p_i \in [0,1]$ is the marginal probability of sub-population $i$ being infected at time $t$ such that $p_i(t) = 1$ if the sub-population $i$ is infected and $p_i (t) = 0$  if it is susceptible. Assuming that every sub-population is experiencing a delay $\tau$ due to the reasons explained earlier, the approximated spread dynamics of sub-population $i$ can then be described using the {\it mean-field approximation} model with time-delay as below, 
\begin{small}
 \begin{align} \label{eq:meta_delay}
    \dot p_i(t)&=-\delta_{i} p_i(t-\tau) + \beta_i \sum_{j=1}^n a_{i j} p_j(t-\tau) \big(1 - p_i(t-\tau) \big) + {\mathbf b}_i^{\rm T}{\boldsymbol \xi}(t),
 \end{align}
 \end{small}
where $\beta_{i}$ is the infection rate at which sub-population $j$ will contaminate sub-population $i$ and also the rate in which the infected individuals in sub-population $i$ will contaminate its susceptible individuals. $\delta_i$ is the recovery rate of sub-population $i$. $\boldsymbol{\xi} = [\xi_1, \ldots, \xi_l]^{\rm T}$ is the effect of an uncertainty in the disease spread dynamics modeled as the vector of independent Gaussian white noise with zero mean and ${\mathbf b}$ as the uncertainty coefficient vector. $a_{i j}$ is the $ij^{th}$ component of the adjacency matrix of the coupling graph. The compact form of meta-population SIS model can be expressed as below,
\begin{align}
\label{eq:system0}
\dot {\mathbf{p}}(t) &= \mathcal{A} \mathbf{p}(t-\tau) - P(t-\tau) B A \mathbf{p}(t-\tau)+\mathcal{B} \boldsymbol{\xi}(t), 
\end{align}
where $P(t-\tau)=\diag \left(\mathbf{p}(t-\tau)\right)$, and infection and recovery matrices are
\begin{align}
    B &~=~\diag \left([\beta_1, \ldots, \beta_n] \right),~\text{and}\nonumber \\
    \Delta &~=~\diag \left([\delta_1, \ldots, \delta_n] \right).
    \label{eq:BD-matrices}
\end{align}

Matrix $\mathcal{A}= B A - \Delta$ determines the stability and performance of network dynamics. 
\begin{equation} \label{eq:mathA}
     \mathcal{A} = \\
    \begin{bmatrix}
   \beta_1 a_{11} -\delta_{1} & \beta_1 a_{1 2} & \cdots & \beta_1 a_{1 n}\\
    \beta_2 a_{2 1} & \beta_2 a_{22}-\delta_{2} & \cdots & \beta_2 a_{2 n}\\
    \vdots & \vdots & \ddots & \vdots\\
    \beta_n a_{n 1} & \beta_n a_{n 2} & \cdots & \beta_n a_{nn}-\delta_{n}
    \end{bmatrix}. 
\end{equation}

In spread dynamics~\eqref{eq:meta_delay}, $\mathcal{B} = [\mathbf{b}_1, \ldots, \mathbf{b}_n]^{\rm T} \in \mathbb{R}^{n\times l}$ is the noise coefficient matrix and $p_i(t)$ is the infected {\em proportion} of sub-population $i$ at time $t$. For every sub-population in the selected community, it can be assumed that $p_i(t)$ is a value close to zero. For instance, this is a valid assumption for the recent COVID-19 pandemic, because even though there were a substantial number of infected individuals from the beginning, the proportion of the infected population between January 19 (The day that first case was reported) through February 28 within the US was still close to zero\footnote{\url{https://www.cdc.gov/coronavirus/2019-ncov/cases-updates/cases-in-us.html}}. It therefore makes sense to linearize the epidemic model around the zero state. Assuming $p_i(t) \ll 1$, equation \eqref{eq:meta_delay} can be linearized as follows:
\begin{equation}
    \dot p_i(t) = -\delta_{i} p_i(t-\tau) + \beta_i \sum_{j=1}^n a_{i j} p_j(t-\tau) + {\mathbf b}_i^{\rm T}{\boldsymbol \xi}(t).
\end{equation}

The compact version of this network is in the following form. 
\begin{align}
\label{eq:system}
\dot {\mathbf{p}}(t)~&= \mathcal{A} \mathbf{p}(t-\tau)+\mathcal{B} \boldsymbol{\xi}(t), 
\end{align}

It should be noted that the following assumptions apply throughout the rest of this paper. 
\begin{assumption}
The infection rate of all the sub-populations is equal. Therefore, $\beta_i = \beta$ for all $i \in \VV$.
\end{assumption} 
\begin{assumption}
The network graph is assumed to be undirected, i.e., $a_{i j} = a_{j i}$.
\end{assumption}
\subsection{Stability analysis}
For the undirected network \eqref{eq:system0}, the initial infection $\mathbf{p}(0) \in [0,1]^n$ will exponentially die out if
\begin{equation} \label{eq:cond1}
    \lambda_{\max} (\mathcal A) \leq -\epsilon.
\end{equation}
where $\lambda_{\max}(.)$ provides the maximum real eigenvalue and $\epsilon \rightarrow 0^+$. In other words, an $\alpha > 0$ could be found to satisfy $\| p_i(t)\| \leq \alpha \|p_i(0)\| e^{-\epsilon t}$ and as a result, the disease-free equilibrium of the system is globally exponentially stable with rate $\epsilon$ \cite{preciado2013}.

The minimum eigenvalue of $\mathcal{A}$ is also lower-bounded by the time-delay as below,
\begin{equation} \label{eq:cond2}
    \lambda_{\min}(\mathcal{A}) \geq -\frac{\pi}{2 \tau}
\end{equation}
which is the direct result of frequency domain stability analysis of the delayed systems \cite{Olfati:2004}. The network is globally asymptotically stable if and only if this condition is satisfied. Therefore, the stability of the network depends not only on the maximum but also the minimum eigenvalue of $\mathcal{A}$.

Combining conditions \eqref{eq:cond1} and \eqref{eq:cond2} results in the following matrix inequality with respect to the positive semi-definite cone $\mathbb{S}^n_+$,

\begin{equation}
    \epsilon I_n \preceq -\mathcal{A} \preceq \frac{\pi}{2 \tau}I_n.
\end{equation}

The reproductive characteristics of a disease determine if it will result in an epidemic or not. The basic reproduction number, $\mathcal{R}_{0M}$, of the epidemic meta-population \eqref{eq:system0} can be defined as below,
\begin{equation}
    \mathcal{R}_{0M} := \frac{\beta}{\delta} \lambda_{\max}\left(A\right)
\end{equation}

The initial infection will converge to zero if $\mathcal{R}_{0M} < 1$ \cite{preciado2013optimal}.
The reproduction number of every sub-population $i$ on the other hand, determines the progress of disease within a single population,
\begin{equation}
    \mathcal{R}_{0S} (i):= \frac{\beta_i a_{ii}}{\delta_i}
\end{equation}
where for $\mathcal{R}_{0S} <1$ the infection size will converge to zero. 
\subsection{Performance analysis}
The steady-state performance, $\rhoo$, of network \eqref{eq:system} can be expressed as,
\begin{equation}
    \rhoo(\mathcal{A} ;\mathcal{B}; \tau)=\lim _{t \rightarrow \infty} \mathbb{E}\left[\mathbf{y}(t)^{\mathrm{T}} \mathbf{y}(t)\right]
\end{equation}
where $\mathbf{y}(t) =
\left[p_1 (t),\ldots,p_n (t)\right]^{\rm{T}}$ is the vector of infection probability at time $t$ which is required to monitor and update the network performance measure. $\mathbf{y}(t)$ is a function of $\mathcal{A}$; therefore, the bounds on the eigenvalues of $\mathcal{A}$ determine the range in which the long-run performance of the network will change.
More details on the performance measure of a class of consensus networks under the influence of exogenous white noises can be found in the reference papers \cite{bamieh2012coherence,young2010robustness,siami2016fundamental, ghaedsharaf2018performance}. 
According to \cite{Doyle89}, the performance of a network can also be found by the frequency domain definition of its $\HH_2$-norm as below,
\begin{align}\label{eq:H2normCalc0}
	\rhoo(\mathcal{A} ;\mathcal{B}; \tau)=\frac{1}{2 \pi} \operatorname{Tr}\left[\int_{-\infty}^{+\infty} G^{\mathrm{H}}(j \omega) G(j \omega) d \omega\right],
\end{align}
where $G(j \omega)$ is the transfer function of the network and $G^{\mathrm{H}}(j \omega) $ corresponds to the complex conjugate transpose of $G(j \omega)$.

\begin{lemma}
For the undirected network \eqref{eq:system}, the closed form solution of \eqref{eq:H2normCalc0} is,
\begin{equation} 
	\rhoo(\mathcal{A} ;\mathcal B; \tau)=\sum_{i=1}^{n} -\frac{\Phi_{i}}{2 \lambda_{i}} \frac{\cos \left(\lambda_{i} \tau\right)}{1+\sin \left(\lambda_{i} \tau\right)},
	\label{eq:perf-meas}
\end{equation}
in which $\Phi_{i}$ is the $i^{th}$ diagonal element of the matrix  $Q^{\T} \mathcal{B} \mathcal{B}^{\mathrm{T}} Q$, where $Q=[\mathbf{q}_1, \dots , \mathbf{q}_{\N}] \in \mathbb{R}^{n\times n}$ is the orthonormal matrix of eigenvectors of $\mathcal{A}$. $\lambda_{i}\left(\mathcal{A}\right)$ for $i=1,2,\ldots,n$ is the $i^{th}$ eigenvalue of matrix $\mathcal{A}$. 
\end{lemma}
\begin{pf}
The transfer matrix of \eqref{eq:system} is,
\begin{align} \label{eq:tf0}
	G(s) &= \left(s I_{n}-e^{-\tau s} \mathcal{A} \right)^{-1} \mathcal{B} \nonumber\\
	& =Q \operatorname{diag}\left(\left[\frac{1}{s-\lambda_{1} e^{-\tau s}}, \cdots, \frac{1}{s-\lambda_{n} e^{-\tau s}}\right]^{\mathrm{T}}\right) Q^{\mathrm{T}} \mathcal{B}.
\end{align}

For this transfer function matrix we have,
\begin{small}
\begin{align}
\label{eq:GHG0}
    & \operatorname{Tr}\left[G^{\mathrm{H}}(j \omega) G(j \omega)\right]= \nonumber\\
    & \operatorname{Tr}\left[Q^{\mathrm{T}} \mathcal{B} \mathcal{B}^{\mathrm{T}} Q \operatorname{diag}\left(\left[\frac{1}{-\lambda_{1} e^{j \tau \omega}-j \omega}, \cdots, \frac{1}{-\lambda_{n} e^{j \tau \omega}-j \omega}\right]^{\mathrm{T}}\right)\right] \nonumber\\
	&\quad\left[\operatorname{diag}\left(\left[\frac{1}{j \omega-\lambda_{1} e^{-j \tau \omega}}, \cdots, \frac{1}{j \omega-\lambda_{n} e^{-j \tau \omega}}\right]^{\mathrm{T}}\right)\right]
\end{align}
\end{small}
and by substituting \eqref{eq:GHG0} in \eqref{eq:H2normCalc0}, the performance will be,
\begin{align} 
\rhoo(\mathcal{A} ;\mathcal{B}; \tau)=\frac{1}{2 \pi} \sum_{i=1}^{n} \int_{-\infty}^{+\infty} \frac{\Phi_{i} d \omega}{\left(j \omega+\lambda_{i} e^{j \tau \omega}\right)\left(\lambda_{i} e^{-j \tau \omega}-j \omega\right)}.
\end{align}

A proof follows by simplifying the above integral.
\end{pf}

It should be noted that the smaller values of $\rhoo$ result in a better performance, therefore, a lower value of performance is desired. 
\section{Centrality indices} \label{sec3}
The importance of every sub-population 
in the disease spread can be analysed by various indices. In this study, the centrality index, $\eta_i$, is the basis to rank the sub-populations. 

For network \eqref{eq:system}, let ${\xi_i(t) \sim \mathcal{N}(0,\sigma_i^2)}$ be the noise affecting the $i^{th}$ sub-population's infection dynamics, which might stem from modeling imperfections, testing error or inaccurate epidemic rates. 
The centrality of sub-population $i$ 
is then defined as the rate of performance with respect to disturbance variance,
 \begin{align}\label{eq:def1}
	\eta_i &\coloneqq \frac{\partial \rhoo}{\partial \sigma_i^2}. 
 \end{align}

The centrality index with respect to two important sources of disturbance will be established in the following sections. 
\subsection{Modeling error}
Model simplifications implemented on the epidemic dynamics affect the state of the infected sub-populations as below,
\begin{equation} \label{eq:ComError}
    \dot p_i(t) = -\delta_{i} p_i(t-\tau) + \beta_{i} \sum_{j=1}^n a_{i j} p_j(t-\tau) + \xi_{i}(t),
\end{equation}
where $\xi_{i}(t) = \sigma_{i} \hat{\xi_{i}}$. The compact form of \eqref{eq:ComError} would be, 
\begin{align}
\label{eq:system1}
\dot {\mathbf{p}}(t) = \mathcal{A} \mathbf{p}(t-\tau)+ \mathcal{B}_1 \hat{\boldsymbol{\xi}}(t), 
\end{align}
in which $\mathcal{B}_1 = \operatorname{diag} \left( \left[ \sigma_1, \ldots, \sigma_{n} \right]\right) \in \mathbb{R}^{n\times n}$.
\begin{corollary}
For the network \eqref{eq:system1}, the centrality index of the $i^{th}$ sub-population is,
\begin{align*}
	\eta_{i}(\mathcal{A}; \tau)=-\frac{1}{2} \left[\mathcal{A}^{-1} \cos (\tau \mathcal{A})\left(I_{n}+\sin (\tau \mathcal{A})\right)^{-1}\right]_{i i},
\end{align*}
for all $i \in {\V}$.
\end{corollary}
\begin{pf}
In the case of having modeling noise, the network dynamics is the same as \eqref{eq:system} with $\mathcal{B} = \mathcal{B}_1$. Hence, the performance will be in the following matrix operator form,
\begin{small}
\begin{align} \label{eq:rhoss}
	&\rhoo(\mathcal{A} ;\mathcal{B}_1; \tau) = \nonumber \\ &-\frac{1}{2} \operatorname{Tr}\left[\operatorname{diag}\left(\left[\sigma_{1}^{2}, \ldots, \sigma_{n}^{2}\right]^{\mathrm{T}}\right) \mathcal{A}^{-1} \cos (\tau \mathcal{A})\left(I_{n}+\sin (\tau \mathcal{A})\right)^{-1} \right].
\end{align}
\end{small}
On the other hand, the value of centrality measure $\rhoo$ is a linear function of variance of elements of noise input. For the centrality index \eqref{eq:def1} the performance is defined as below, 
\begin{equation} \label{eq:rhodef-node}
    \rhoo=\sum_{i \in \VV} \eta_{i} \sigma_{i}^{2}.
\end{equation}

Substituting equation \eqref{eq:rhoss} in the above definition, the centrality will be obtained.
\end{pf}

\subsection{Testing error}
In many cases of epidemic, especially when an infectious disease like COVID-19 first emerges, the testing methods are not completely accurate in terms of identifying the infected individuals. The incorrect results generate inaccurate statistics regarding the population of the confirmed cases which triggers impaired judgment and inappropriate containment methods.  

In theory, the testing error affects every sub-population's state in the following way,
\begin{small}
\begin{equation}
\label{eq:agentUpdateSN}
\dot p_i(t)  = -\delta_{i} \left(p_i(t-\tau)+\xi_{i}(t)\right) + \beta_{i} \sum_{j=1}^n  a_{i j} \left(p_j(t-\tau)+\xi_{j}(t) \right),
\end{equation}
\end{small}
where $\xi_i \sim \mathcal N (0, \sigma_i^2)$ for $i \in \V$. The state of the infected population in this case will be the same as \eqref{eq:system} with $\mathcal{B} = \mathcal{B}_2= \mathcal{A}\operatorname{diag} \left( \left[ \sigma_1, \ldots, \sigma_{n} \right]\right) \in \mathbb{R}^{n\times n} $ as below,
\begin{align}
\label{eq:system2}
\dot {\mathbf{p}}(t) = \mathcal{A} \mathbf{p}(t-\tau)+ \mathcal{B}_2 \hat{\boldsymbol{\xi}}(t).
\end{align}
\begin{corollary}
For the network \eqref{eq:system2}, the centrality index is,
\begin{align}\label{etaiSensor}
	\eta_{i}(\mathcal{A}; \tau)=-\frac{1}{2}\left[\mathcal{A} \cos (\tau \mathcal{A})\left(I_{n}+\sin (\tau \mathcal{A})\right)^{-1} \right]_{i i},
\end{align}
for all $i \in \V$.
\end{corollary}
\begin{pf}
With the testing error noise, the network dynamics is the same as \eqref{eq:system} with $\mathcal{B} = \mathcal{B}_2$ . Therefore, the compact form of $\rhoo$ is,
\begin{align}
	&\rhoo(\mathcal{A} ;\mathcal{B}_2; \tau) = \nonumber \\ &\frac{1}{2} \operatorname{Tr}\left[-\operatorname{diag}\left(\left[\sigma_{1}^{2}, \ldots, \sigma_{n}^{2}\right]^{\mathrm{T}}\right)  \mathcal{A} \cos (\tau \mathcal{A})\left(I_{n}+\sin (\tau \mathcal{A})\right)^{-1}  \right].
\end{align}

A proof follows by using the definition \eqref{eq:rhodef-node} and above equation to find centrality.
\end{pf}

\section{Epidemic containment by traffic volume optimization at community levels} \label{sec4}

\subsection{Optimal traffic restriction}
Monitoring and regulation of the traffic volume is one of the potential government interventions to mitigate the epidemic threat. Regarding the underlying epidemiological network, traffic restriction between two sub-populations directly changes the corresponding edge weight, $w_e$, of those sub-populations in the network. Therefore, the stability around disease-free state could be obtained by monitoring and management of the transportation network and restriction of the traffic volume between the highly infected and highly susceptible candidates. Although the complete isolation of the sub-populations seems to be the easiest and safest prevention method, especially in the case of COVID-19 which has now lasted for several months, it is not a permanent solution mostly because of the economic considerations. Therefore, a proper balance must be found in the decision-making process. In this study, a convex optimization method is offered to determine the proper volume of the transportation by minimizing the value of corresponding network performance, and consequently improving the performance, with respect to the graph weights.

This optimal traffic control problem for the network with modeling error noise can be expressed as below,
\begin{align} \label{prob:opt1}
		 \underset{{w_e,\forall e\in \EE}}{\text{minimize }}\hspace{0.4cm}
		&\hspace{1cm}   \rhoo(\mathcal{A}; \mathcal{B}_1;\tau)  \\
		 \text{subject to:\hspace{0.4cm}}
		&  \underline{w_e} \leq w_e \leq \overline{w_e} \nonumber\\
		&  \sum_{e \in \EE} w_e ~=~ c \nonumber \\
		& \mathcal{A} ~=~ B \sum_{e \in \mathcal E}  w_e A_e + B \diag \left([a_{11}, \ldots, a_{nn}] \right) - \Delta \nonumber \\
		& - \frac{\pi}{2 \tau} I_n ~\preceq~ \mathcal{A}~\preceq~ -\epsilon I_n. \nonumber
\end{align}

Here, the first constraint determines lower and upper bounds for every the network weight. $\underline{w_e}$ and $\overline{w_e}$ are the minimum and maximum traffic flow in edge $e$ which are arbitrary values selected by economic considerations or traffic management methods to avoid decreasing traffic volume less than a minimum threshold or increasing it more than the maximum capacity. The second constraint determines the total weight of the network edges or overall traffic volume, which could acquire any desired value $c$ depending on the intensity of isolation. The third constraint in which $A_e$ is the adjacency matrix of link $e$ gives the definition of $\mathcal{A}$ with respect to edge weights. The last constraint imposes another limitation on the edge weights to respect the domain of stable solutions. $B$ and $\Delta$ are the infection and recovery rate matrices defined in equation \eqref{eq:BD-matrices}.
Consider an optimization problem with the following general form,
\begin{align}
		 \underset{x}{\text{minimize }}\hspace{0.4cm}
		&\hspace{1cm}   f_0(x) \label{prob:opt0} \\
		 \text{subject to:\hspace{0.4cm}}
		&  f_i(x) \leq b_i, \quad i=1, \ldots, q \nonumber \\
		&  h_{i}(x)=0, \quad i=1, \ldots, r. \nonumber
\end{align}

This problem is considered a convex problem if all the functions $f_0,\ldots,f_q$ are convex and all the equality constraints $h_1,\ldots,h_r$ are affine \cite{boyd2004convex}. According to this definition, the problem \eqref{prob:opt1} does not fall into the category of convex problems. Hence, some modifications need to be implemented on the original optimal problem \eqref{prob:opt0}. In this regard, an approximation of the performance has been offered by \cite{ghaedsharaf2018performance} which converts the product of non-convex trigonometric functions to a linear function of $\mathcal{A}$ and its inverse. Using this approximation, the network performance will be,
\begin{align}
    \rhoo(\mathcal{A}; \mathcal{B}_1;\tau) \simeq & \frac{1}{2} \operatorname{Tr}\left[-\mathcal{A}_{o} \mathcal{A}^{-1}+\frac{4 \tau}{\pi} \mathcal{A}_{o}\left(\frac{\pi}{2} I_{n}+\tau \mathcal{A}\right)^{-1} \right. \nonumber \\ 
    & \left. -c_{1} \tau^{2} \mathcal{A}_{o} \mathcal{A}+\frac{c_{0}}{2} \tau \mathcal{A}_{o}\right],
\end{align}
where $\mathcal{A}_{o} = \mathcal{B}_1 \mathcal{B}_1^{\rm T}$ and the constant parameters $c_0 = 0.1873$ and $c_1=-0.01$ are estimated to minimize the mean squared error of the approximated performance. This approximation is still not a convex function, as it includes the non-convex inverse functions $\mathcal{A}^{-1}$ and $\left(\frac{\pi}{2} I_{n}+\tau \mathcal{A}\right)^{-1}$. Substituting the epigraph variables $X_1 = \mathcal{A}^{-1}$ and $X_2 = \left(\frac{\pi}{2} I_{n}+\tau \mathcal{A}\right)^{-1}$ will turn a convex function. The optimization problem \eqref{prob:opt1} can now be cast as the following approximate form, which is a convex optimization problem as the objective function as well as all the inequality constraints are convex and the equality constraints are affine with respect to the only optimization variable $w_e$.

\begin{align}
		 \underset{X_1;X_2;w_e,\forall e\in \EE}{\text{minimize }}\hspace{0.0cm}
		&\hspace{1cm}   \operatorname{Tr}\left[\mathcal{A}_{o} X_1+\frac{4 \tau}{\pi} \mathcal{A}_{o} X_2-c_{1} \tau^{2} \mathcal{A}_{o} \mathcal{A}\right] \label{prob:opt1-1}\\
		 \text{subject to:\hspace{0.4cm}}
		&  \underline{w_e} \leq w_e \leq \overline{w_e}\nonumber\\
		&  \sum_{e \in \EE} w_e ~=~ c \nonumber \\
		& \mathcal{A} ~=~ B \sum_{e \in \mathcal E}  w_e A_e + B \diag \left([a_{11}, \ldots, a_{nn}] \right) - \Delta \nonumber\\
		& \mathcal{A} + \frac{\pi}{2 \tau} I_n \succeq 0 \nonumber\\
		& - \mathcal{A} -\epsilon I_n \succeq 0 \nonumber\\
		&\left[\begin{array}{cc}
		X_{1} & I_n \\ I_n & -\mathcal{A} \end{array}\right] \succeq 0 \nonumber\\
		& \left[\begin{array}{cc}
		X_{2} & I_n \\ I_n & \frac{\pi}{2} I_{n}+\tau\mathcal{A}
		\end{array}\right] \succeq 0.	\nonumber
\end{align}
\useshortskip
\indent
\subsection{Robust traffic restriction}
While the target of proposed optimal approach is to improve the overall network performance with a uniform uncertainty distribution, i.e. $\sigma_i=1$ for $i=1,\ldots,n$, there are cases in which the sub-populations with highest centrality experience higher levels of disturbance. Worst case noise distribution highlights the role of high centrality sub-populations in epidemic growth which requires us to design a robust containment approach. 

In this section, the traffic restriction problem is investigated as a robust design optimization where the worst-case scenario is the optimization target. Such a case can be expressed as a min-max optimization presented in problem \ref{prob:opt2}. The first constraint limits the sum of squared noise variances and the rest of the constraints are the same as explained for problem \eqref{prob:opt1}.  
The inner optimization loop is to find the highest performance of the network with respect to the uncertainty $\sigma_i$ in order to improve the robustness against the disease spread. 
    \begin{flalign}\label{prob:opt2}
 		\begin{aligned}
 		& \underset{w_e,\forall e\in \EE}{\text{minimize}}
 		& & \underset{\sigma_i,\forall i\in \VV}{\text{maximize }} \rhoo(\mathcal{A}; \mathcal{B}_1; \tau) \\
 		& \text{subject to:}
 		& & \sum_{i \in \VV} \sigma_i^2 = n\\
 		& & & \underline{w_e} \leq w_e \leq \overline{w_e}\\
 		& & & \sum_{e \in \EE} w_e ~=~ c\\
 		& & & \mathcal{A} ~=~ B \sum_{e \in \mathcal E}  w_e A_e + B \diag \left([a_{11}, \ldots, a_{nn}] \right)- \Delta\\
 		& & &- \frac{\pi}{2 \tau} I_n ~\preceq~ \mathcal{A}~\preceq~ -\epsilon I_n. 
 		\end{aligned}
 	\end{flalign}
Using the performance definition in \eqref{eq:rhodef-node}, the inner optimization problem in \eqref{prob:opt2} can be rewritten in the following form,
\begin{align}
		 \underset{\sigma_i,\forall i \in \VV}{\text{minimize }}\hspace{0.0cm}
		&\hspace{.1cm}   \sum_{i \in \VV} \eta_{i}(\mathcal{A}; \tau) \sigma_{i}^{2} \label{prob:opt2-1} \\
		 \text{subject to:\hspace{0.4cm}}
		&  \sum_{i \in \VV} \sigma_i^2 = n. \nonumber
\end{align}
Here, the cost function and constraint are only linear functions of the optimization variable $\sigma_i$. Hence, the maximum performance occurs in the boundary, where for one variable we have $\sigma_i^2 = n$ and $\eta_i(\mathcal{A}; \tau)$ has its maximum value, and for the rest $\sigma_i^2 = 0$, i.e., 
\begin{equation}
\displaystyle \sigma_i^2 := \left\{\begin{array}{cl}
n & \textrm{if}~ i = \argmax_{j\in \mathcal V} \eta_j(\mathcal{A}; \tau) \\
 &   \\
0 & \textrm{otherwise}
\end{array}\right.
\label{worst-case}
\end{equation}
\useshortskip
\indent
As a result, the maximum objective function of problem \eqref{prob:opt2-1} is equal to the maximum value of $\frac{n}{2} \left[-\mathcal{A}^{-1} \cos (\tau \mathcal{A})\left(I_{n}+\sin (\tau \mathcal{A})\right)^{-1}\right]_{i i}$ for all ${i \in \VV}$. Note that this solution is a non-convex function of $w_e$; therefore, to use it in the outer minimization loop, its approximated epigraph version will be used. Problem \eqref{prob:opt2} can now be cast in the following form,
\begin{align}
		 \underset{x; w_e, \forall e \in {\EE}}{\text{minimize }}\hspace{0.4cm}
		&\hspace{1cm}   x 		\label{prob:opt2-3} \\
		 \text{subject to:\hspace{0.4cm}}
		&  \underline{w_e} \leq w_e \leq \overline{w_e} \nonumber\\
		&  \sum_{e \in \EE} w_e ~=~ c \nonumber\\
		& \mathcal{A} ~=~ B \sum_{e \in \mathcal E}  w_e A_e + B \diag \left([a_{11}, \ldots, a_{nn}] \right) - \Delta \nonumber\\
		& \mathcal{A} + \frac{\pi}{2 \tau} I_n \succeq 0 \nonumber\\
		& - \mathcal{A} -\epsilon I_n \succeq 0 \nonumber\\
		& x \succeq \frac{n}{2} \left[ \mathcal{A}_{o} X_1+\frac{4 \tau}{\pi} \mathcal{A}_{o} X_2-c_{1} \tau^{2} \mathcal{A}_{o} \mathcal{A} \right]_{ii},~ \forall i \in \VV \nonumber\\
		&\left[\begin{array}{cc}
		X_{1} & I_n \\ I_n & -\mathcal{A} \end{array}\right] \succeq 0 \nonumber\\
		& \left[\begin{array}{cc}
		X_{2} & I_n \\ I_n & \frac{\pi}{2} I_{n}+\tau\mathcal{A}
		\end{array}\right] \succeq 0. \nonumber 
\end{align}

It is worth mentioning that the same methods in the preceding sections can be implemented on the networks with the described testing error noise as well.
%
\section{Results} \label{sec5}
In this section, the established models and control methods will be implemented on a core-periphery network, as an academic example, and a US airport network, as a real world example. {It should be noted that most of the presented results are based on the assumption that members of every sub-population are not completely following social distancing rules, $a_{ii} \neq 0$ for all $i=1,2,\ldots,n$. Therefore, they are free to join other sub-populations through air transportation. This will allow us to track the disease through the United States.}

\begin{figure*}
    \centering
    \subfigure[]{}
    \includegraphics[width=.345\linewidth,trim={3.5cm 5cm 6cm 0},clip]{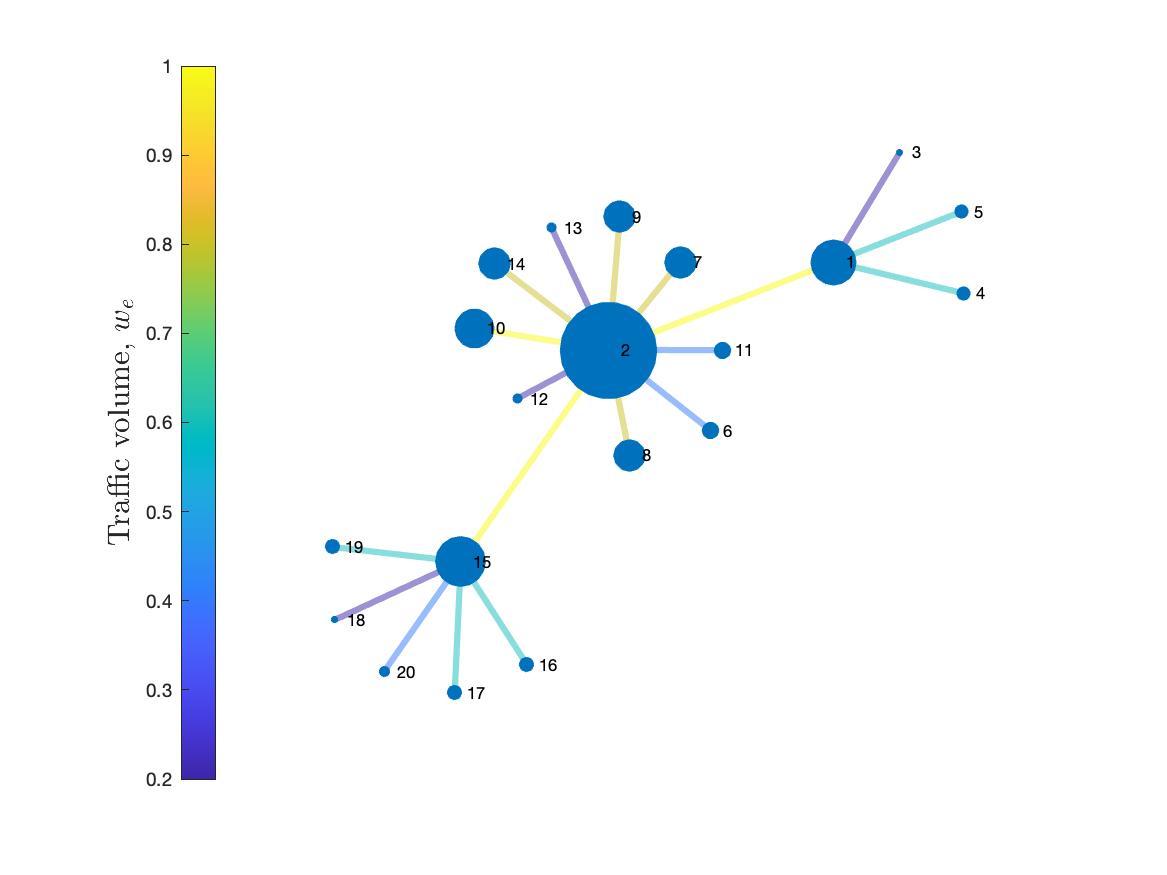}   
    \subfigure[]{}
    \includegraphics[width=.305\linewidth,trim={6cm 5cm 7cm 0},clip]{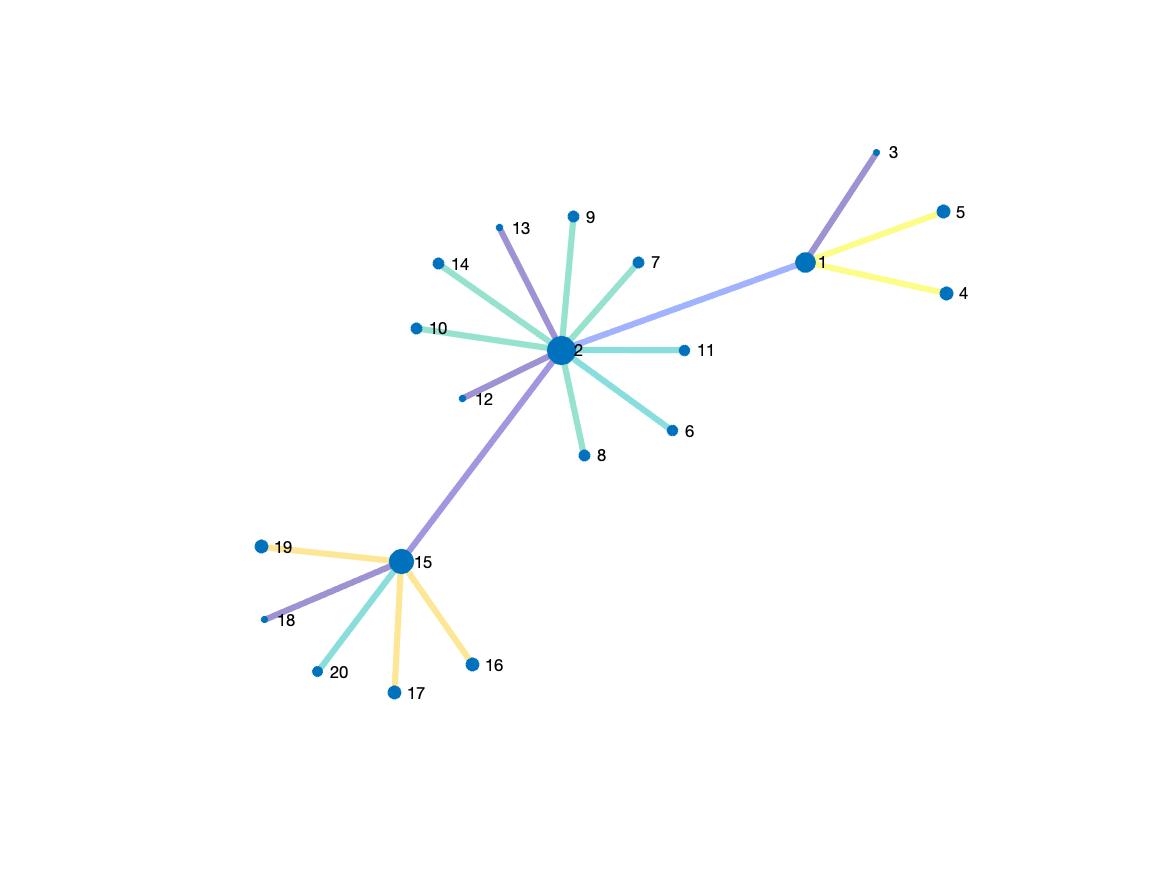}
    \subfigure[]{}
    \includegraphics[width=.305\linewidth,trim={6cm 5cm 7cm 0},clip]{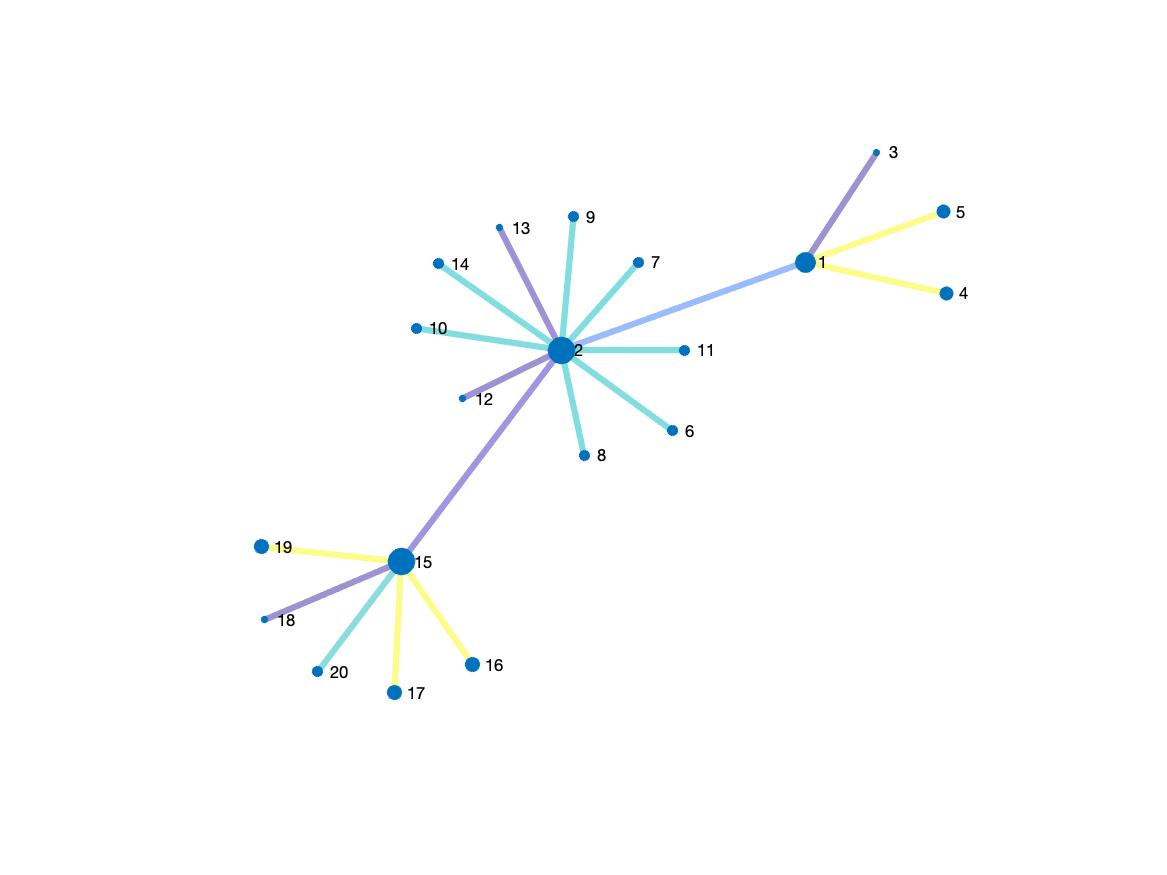}
\caption{(a) Meta-population network of 20 sub-populations and their normal traffic volume. All the sub-populations are experiencing $5$ days lag. The nodes are ranked based on their centrality index, $\eta_i$, which is reflected through the size of their indicating circles. The interconnections are ranked by their corresponding traffic volume which is specified by the color of edges.  (b) Optimal meta-population network of Fig.~\ref{Fig:Tree_Networks}(a) designed by the optimal approach \eqref{prob:opt1-1}, where a uniform noise distribution is applied. (c) Robust meta-population network of Fig.~\ref{Fig:Tree_Networks}(a) designed by the robust approach \eqref{prob:opt2-3}, where the worst case of applying the maximum noise input to the sub-population with highest centrality is considered.}   \label{Fig:Tree_Networks} 
\end{figure*}

\subsection{Core-periphery network}
To evaluate the performance of the proposed methods, a core-periphery network consisting of three communities has been simulated as the representative of a meta-population. The simulations are based on the dynamics \eqref{eq:system0} over a tree graph with three connected star graphs consisting of $20$ nodes and $19$ edges weighted in the range of $[0,1]$ (see Fig.~\ref{Fig:Tree_Networks}(a)). A combination of multiple star graphs is a good candidate for a meta-population, as in reality some of the sub-populations are considered as hubs while the others connect to the rest of the sub-populations through these hubs. In this network, nodes $1$, $2$, and $15$ are the hubs.
In Fig.~\ref{Fig:Tree_Networks}(a), all the sub-populations are experiencing $5$ days of delay and they are ranked based on their centrality index, $\eta_i$, which is reflected through the size of their indicating circles. The interconnections are ranked by their corresponding traffic volume which is specified by the color of edges. Let us assume that the network shown in Fig.~\ref{Fig:Tree_Networks}(a) is infected by a virus and experiencing $5$ days of lag. Using the convex optimization method \eqref{prob:opt1-1} with $c$ being the overall traffic volume in the absence of traffic restrictions, the network structure changes into Fig.~\ref{Fig:Tree_Networks}(b). The result of the robust optimization approach \eqref{prob:opt2-3} for the case of $5$ days of delay is shown in Fig.~\ref{Fig:Tree_Networks}(c). The robust optimization tends to consider the worst case scenario where the uneven noise distribution amplifies the effect of hub node $2$ with the highest centrality, which makes it a bigger threat requiring it to be even more isolated. Hence, the robust optimizer is more conservative in manipulating the traffic volumes and reducing the centrality of hubs.

\begin{figure}
	\centering
	\includegraphics[trim={0 1cm 0 0}, width=0.75\textwidth]{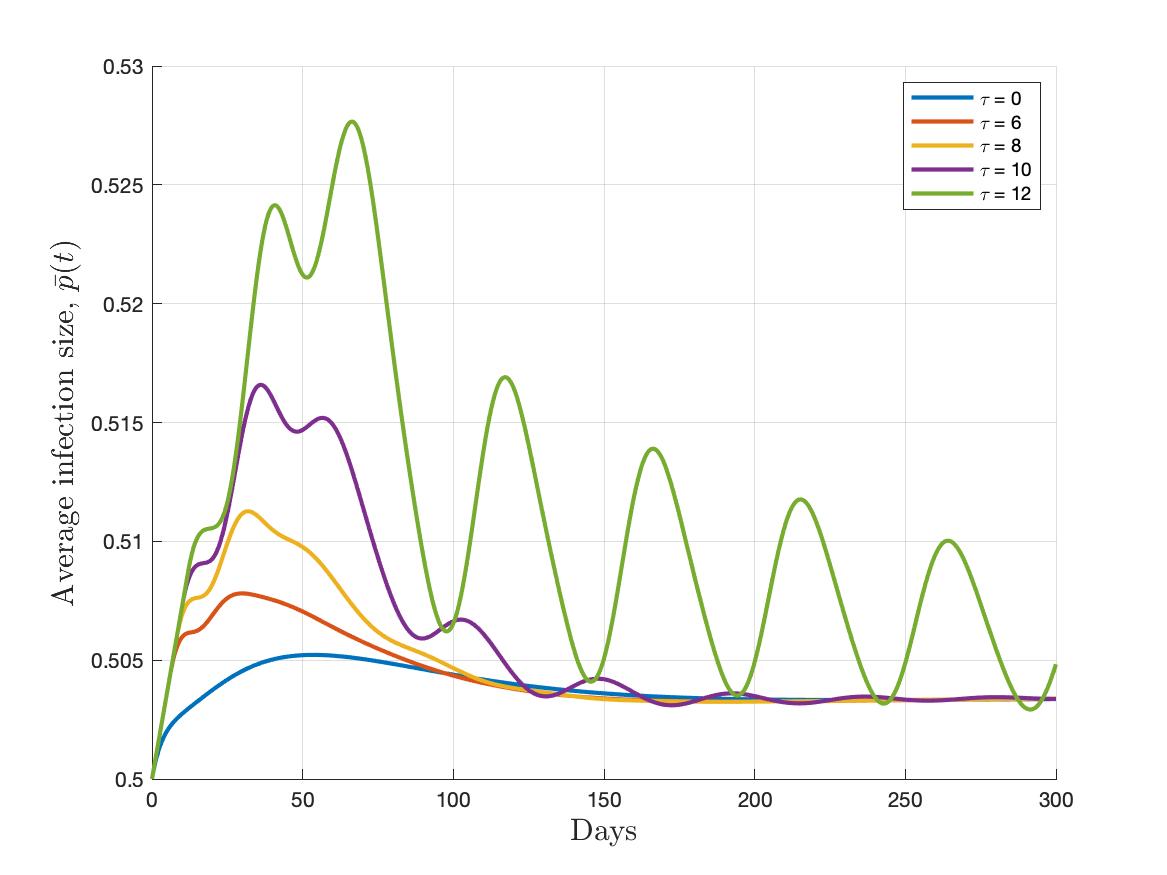}
	\caption{\small{The average infection size, $\bar{p}(t) = \frac{1}{n}\sum_{i \in \mathcal{V}} p_i(t)$, of the meta-population network shown in Fig.~\ref{Fig:Tree_Networks}(a) with different time-delays. $50$ percent of the  meta-population is initially infected and $\mathcal{R}_0=4.7$. It is assumed that there is effective social distancing within every sub-population, $a_{ii}=0$ for $i=1,2,\ldots,n$.}}
	\label{Fig:Tree_Delayeffect}
\end{figure}

The effect of time-delay on the epidemic evolution of the simulated network meta-population when there is effective social distancing within every sub-population ($a_{ii}=0$ for $i=1,2,\ldots,n$) has been illustrated in Fig.~\ref{Fig:Tree_Delayeffect}, where the structure of the modeled network is shown in Fig.~\ref{Fig:Tree_Networks}(a). Based on Fig.~\ref{Fig:Tree_Delayeffect}, it can be concluded that the higher the delays in identifying infected individuals, the higher the risk of experiencing a more extreme epidemic peak with multiple pulses. The time-delay also shows a correlation with the onset of epidemic peak, which is a decisive factor in designing the proper government interventions.

\begin{figure}
	\centering
	\includegraphics[trim={0 1cm 0 0}, width=0.75\textwidth]{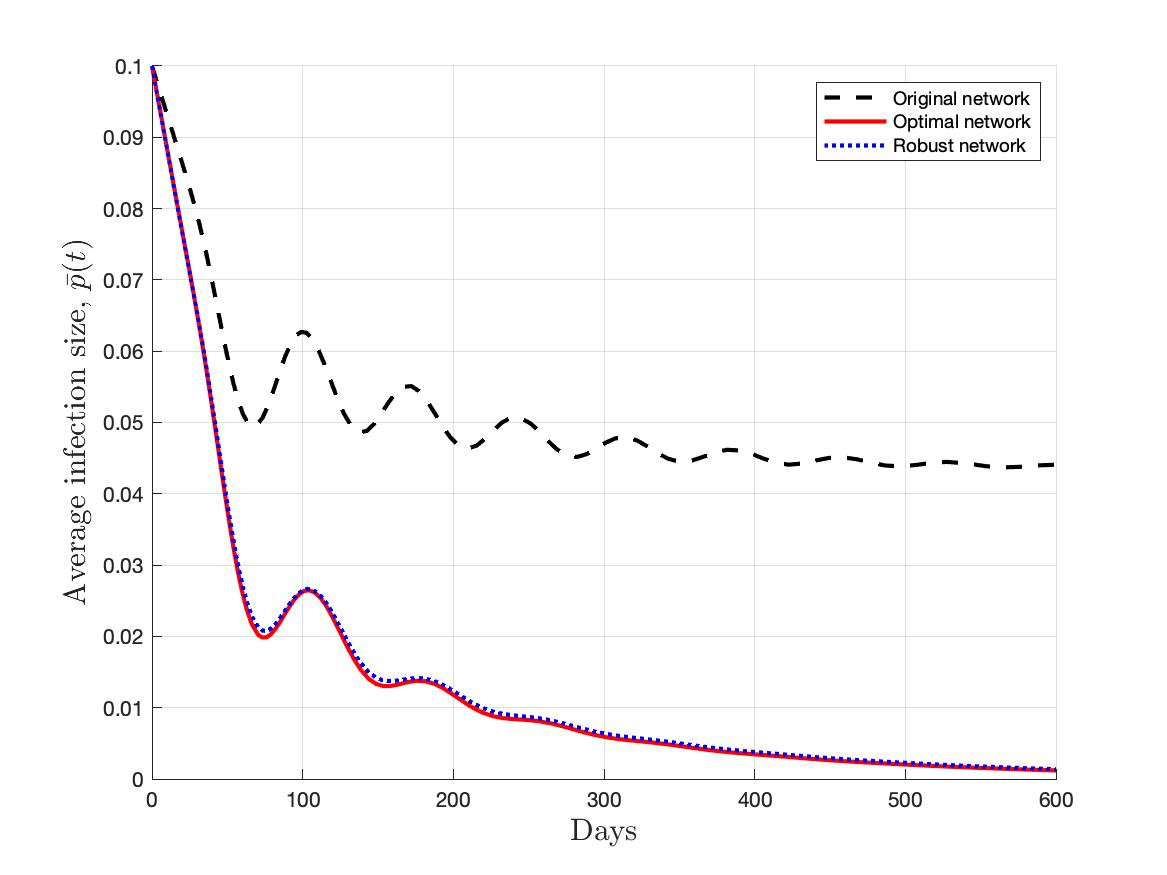}
	\caption{\small{Average infection size of the three networks in Figs.~\ref{Fig:Tree_Networks}(a), \ref{Fig:Tree_Networks}(b), and \ref{Fig:Tree_Networks}(c)with $\tau = 17$ days, $\mathcal{R}_0 \simeq 1.12$, and initial infection of $50$ percent in 4 nodes.}}
	\label{Fig:Tree_Infection}
\end{figure}

The changes in average infection size of network \ref{Fig:Tree_Networks}(a) when the optimal and robust controls are applied is shown in Fig.~\ref{Fig:Tree_Infection}. 
For a meta-population that is initially $10$ percent infected, it is shown that $5$ percent of meta-population will always be infected, while for the controlled network, the infection eventually dies out. 

\begin{figure*}
    \centering
    \subfigure[]{}{
    \includegraphics[width=.31\linewidth,trim={2cm 1cm 2cm 0},clip]{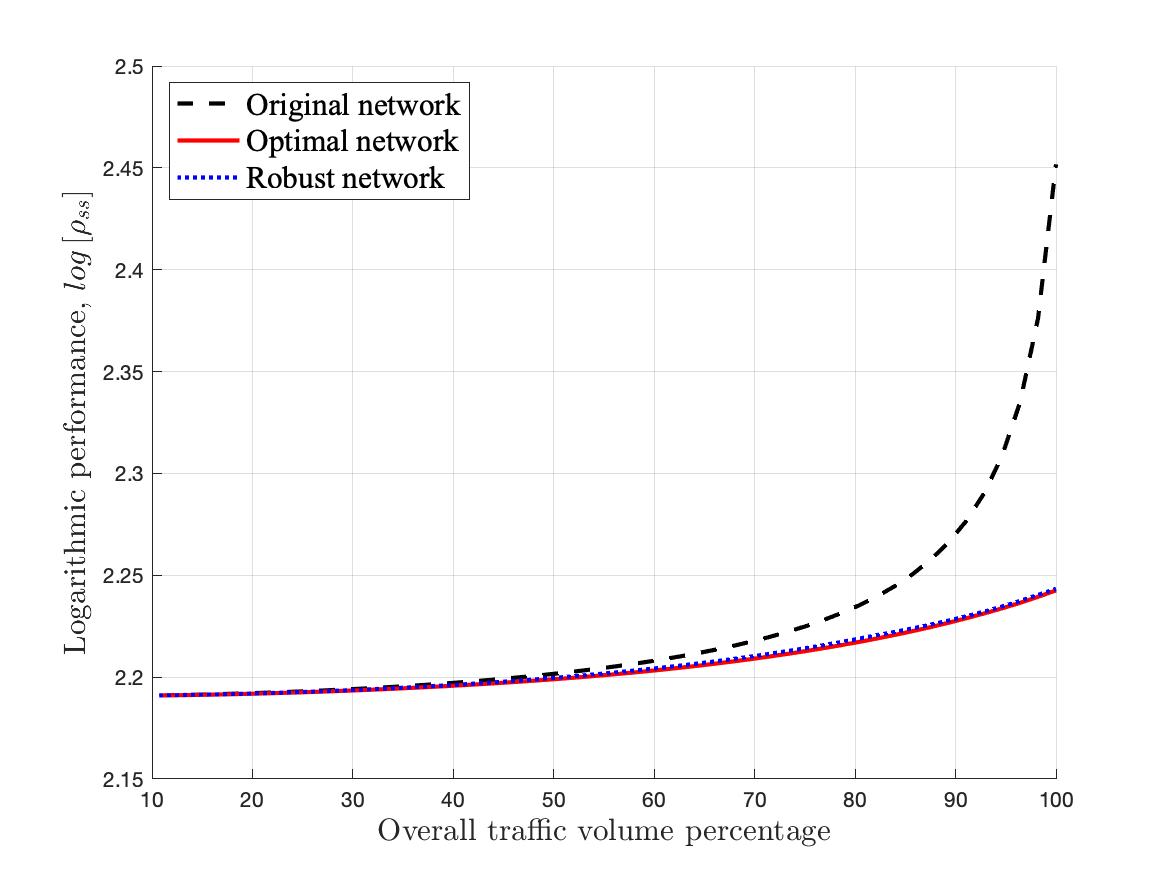}    
       }\hfill
    \subfigure[]{}{
    \includegraphics[width=.31\linewidth,trim={1cm 1cm 3cm 0},clip]{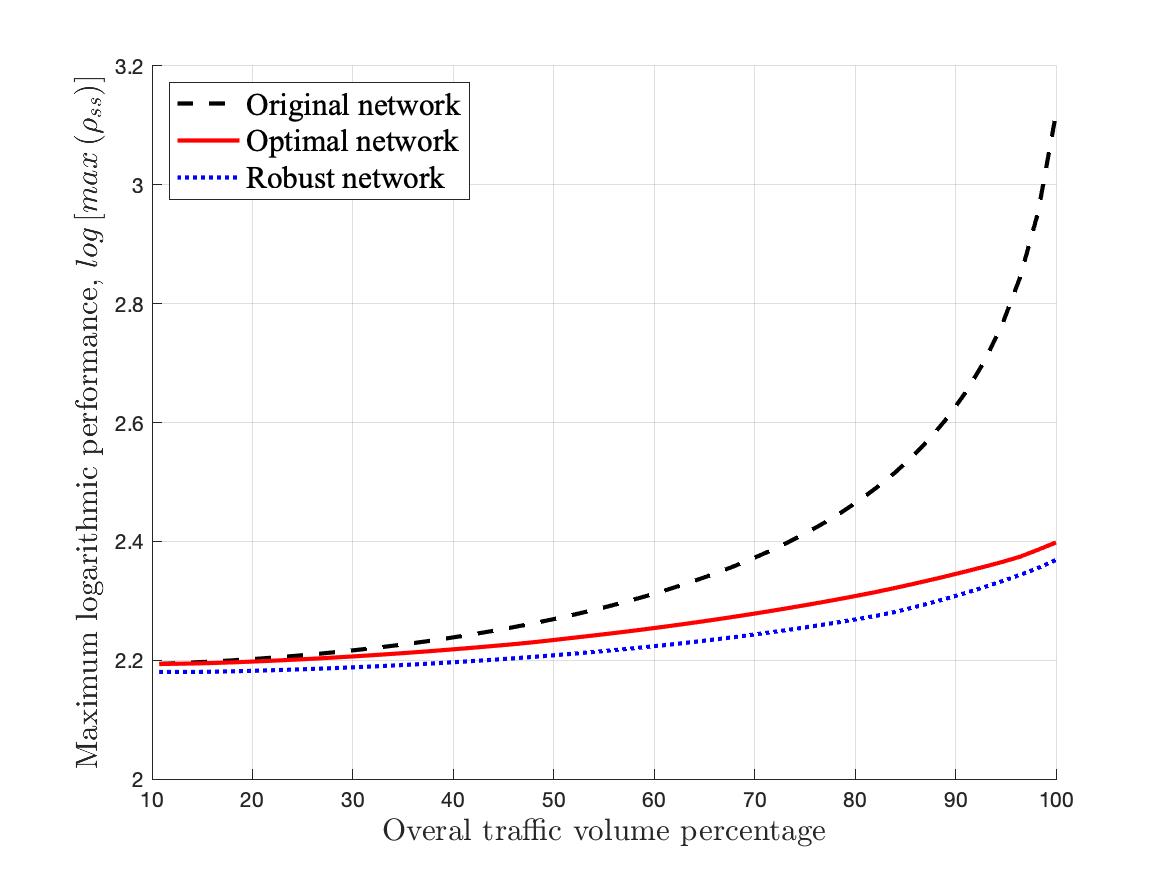}}\hfill
    \subfigure[]{}{
    \includegraphics[width=.31\linewidth,trim={0cm 1cm 3cm 0},clip]{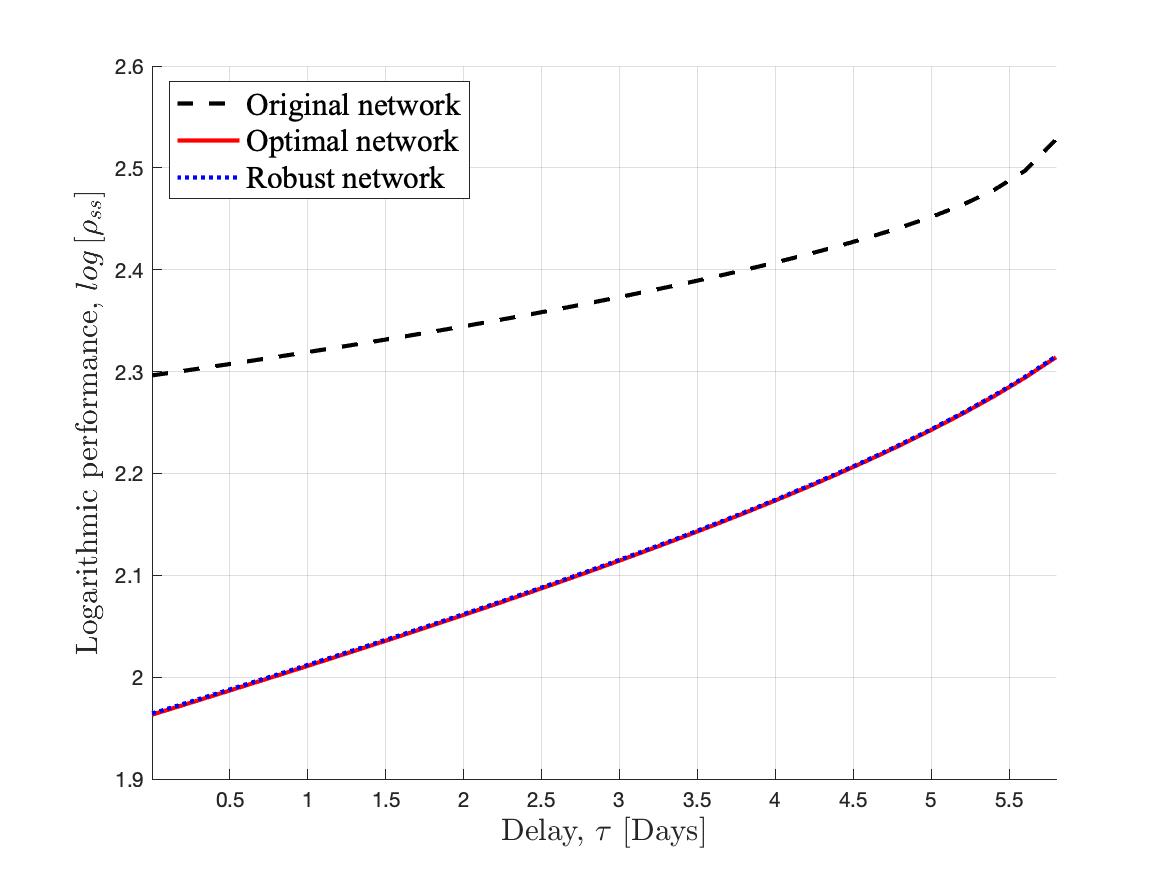}}
\caption{(a) Performance comparison between the original network Fig.~\ref{Fig:Tree_Networks}(a), the optimal network of Fig.~\ref{Fig:Tree_Networks}(b) and the robust network of Fig.~\ref{Fig:Tree_Networks}(c) in the case of uniform noise distribution. (b) Performance comparison between the original network Fig.~\ref{Fig:Tree_Networks}(a), the optimal network of Fig.~\ref{Fig:Tree_Networks}(b) and the robust network of Fig.~\ref{Fig:Tree_Networks}(c) in the case of extreme noise distribution. (c) Logarithmic performance measure of networks Fig.~\ref{Fig:Tree_Networks} with respect to time-delay.}   \label{Fig:Tree_norm} 
\end{figure*}

Fig.~\ref{Fig:Tree_norm}(a) shows the logarithmic scale performance of the original, optimal, and robust networks with respect to the desired traffic volume of the meta-population when all of the sub-populations, regardless of their centrality, experience an equal level of disturbance. As it was mentioned in the preceding sections, the smaller the value of performance, the better the obtained performance. Hence, the optimal and robust controllers have been successful in improving the performance of the original network, while the optimal network results in a slightly better performance due to its less conservative weight distribution. 

Additionally, a comparison between the highest performance of both optimization methods is presented in Fig.~\ref{Fig:Tree_norm}(b) to put more emphasis on the importance of considering worst cases while determining the intensity of traffic restriction during the epidemic. The direct correlation between network performance measure, $\rhoo$, and time-delay can be found in Fig.~\ref{Fig:Tree_norm}(c). As was expected, increasing time-delay impairs the performance.

\begin{figure}
	\centering
	\includegraphics[trim={0 1cm 0 0}, width=0.75\textwidth]{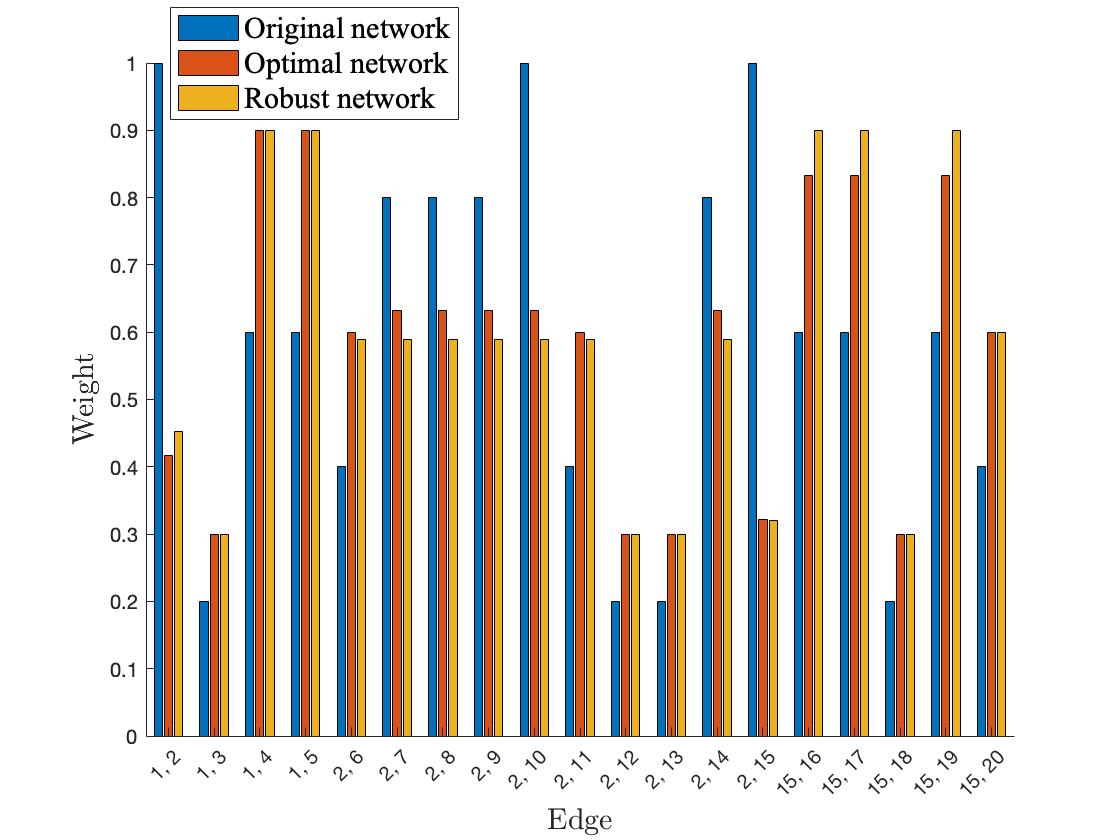}
	\caption{\small{Weight (traffic volume) comparison between the original network Fig.~\ref{Fig:Tree_Networks}(a), the optimal network Fig.~\ref{Fig:Tree_Networks}(b), and the robust network Fig.~\ref{Fig:Tree_Networks}(c).}}
	\label{Fig:Tree_Weightbar}
\end{figure}

\begin{figure}
	\centering
	\includegraphics[trim={0 2cm 0 0}, width=0.75\textwidth]{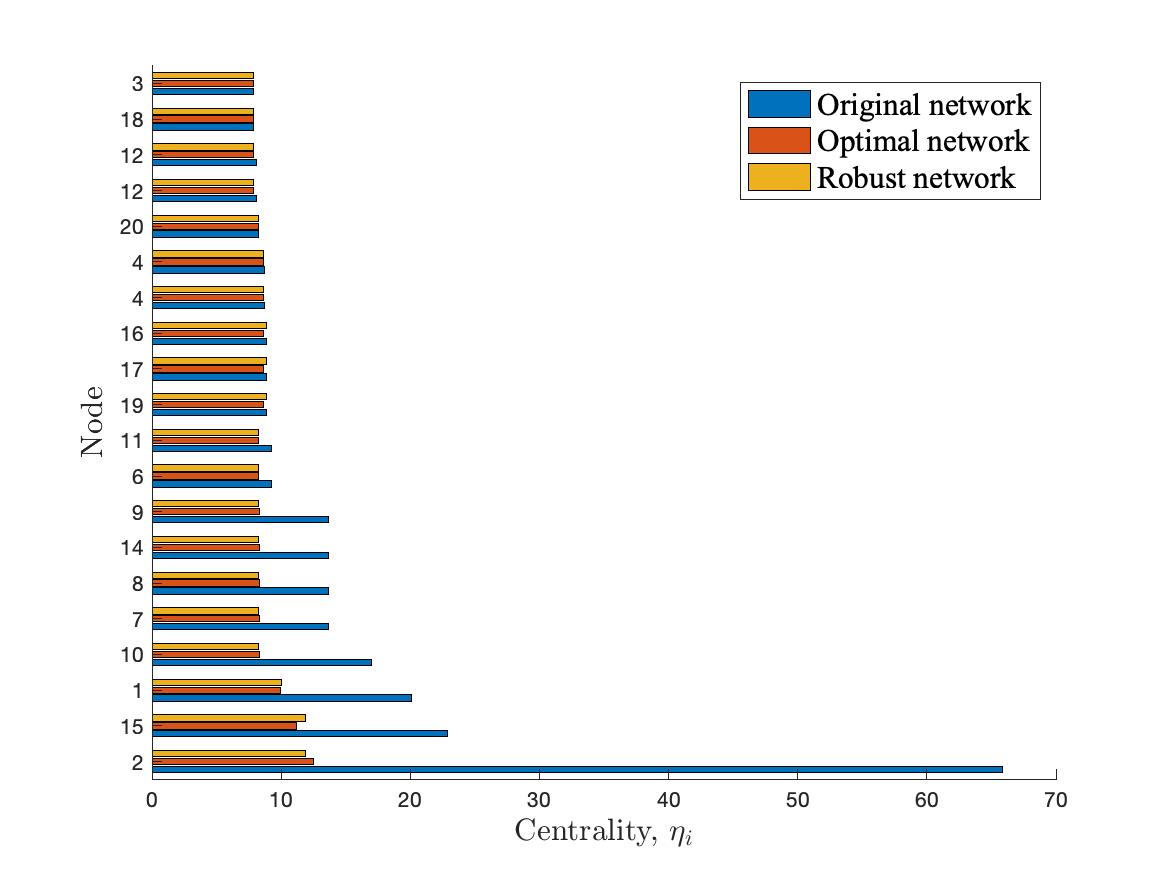}
	\caption{Centrality comparison between the hubs of original network Fig.~\ref{Fig:Tree_Networks}(a), optimal network Fig.~\ref{Fig:Tree_Networks}(b), and robust network Fig.~\ref{Fig:Tree_Networks}(c).}
	\label{Fig:Tree_Centralitybar}
\end{figure}

To make a better comparison between the resulted networks, the bar diagram of the edge weights, or traffic volumes, is represented in Fig.~\ref{Fig:Tree_Weightbar}. It can be seen that between the connections with highest volume, $(1,2)$, $(2,10)$, $(2,11)$, $(2,15)$, and $(15,18)$, only the ones connecting two of three hubs have considerable drop in volume. The centrality diagram of sub-populations can be found in Fig.~\ref{Fig:Tree_Centralitybar}. 

The detailed results of the performance improvement through optimal and robust epidemic controls are represented in Table.~\ref{table_1} for Case $1$ with uniform noise distribution, i.e. $\sigma_i=1$ for $i=1,\ldots,n$, and Case $2$ is the worst case explained in \eqref{worst-case}.
\begin{table}[H]
\small
	\centering
	 \caption{\small{The values and percentages of performance enhancement for the networks shown in Figs. \ref{Fig:Tree_Networks}(b) and \ref{Fig:Tree_Networks}(c) compared with Fig. \ref{Fig:Tree_Networks}(a)}}
	\begin{tabular}{ |c|c|c| } 
\cline{2-3}
    \multicolumn{1}{c|}{} & Case 1 & Case 2\\ \hline 
     Original network (Fig.~\ref{Fig:Tree_Networks}(a))& 283 ~($0\%$)  &  1317 ~($0\%$)\\ 
      \hline
     Optimal network (Fig.~\ref{Fig:Tree_Networks}(b))& 175 ~($+38\%$) & 250 ~($+81\%$)\\ 
      \hline
     Robust network (Fig.~\ref{Fig:Tree_Networks}(c))&  175 ~($+38\%$) & 233 ~($+82\%$)\\ 
     \hline
\end{tabular}
 \label{table_1}
\end{table}

\subsection{Network of United States busiest airports}
\begin{figure*}
    \centering
    \subfigure[]{}
    \includegraphics[width=.34\linewidth,trim={0 8cm 8cm 0},clip]{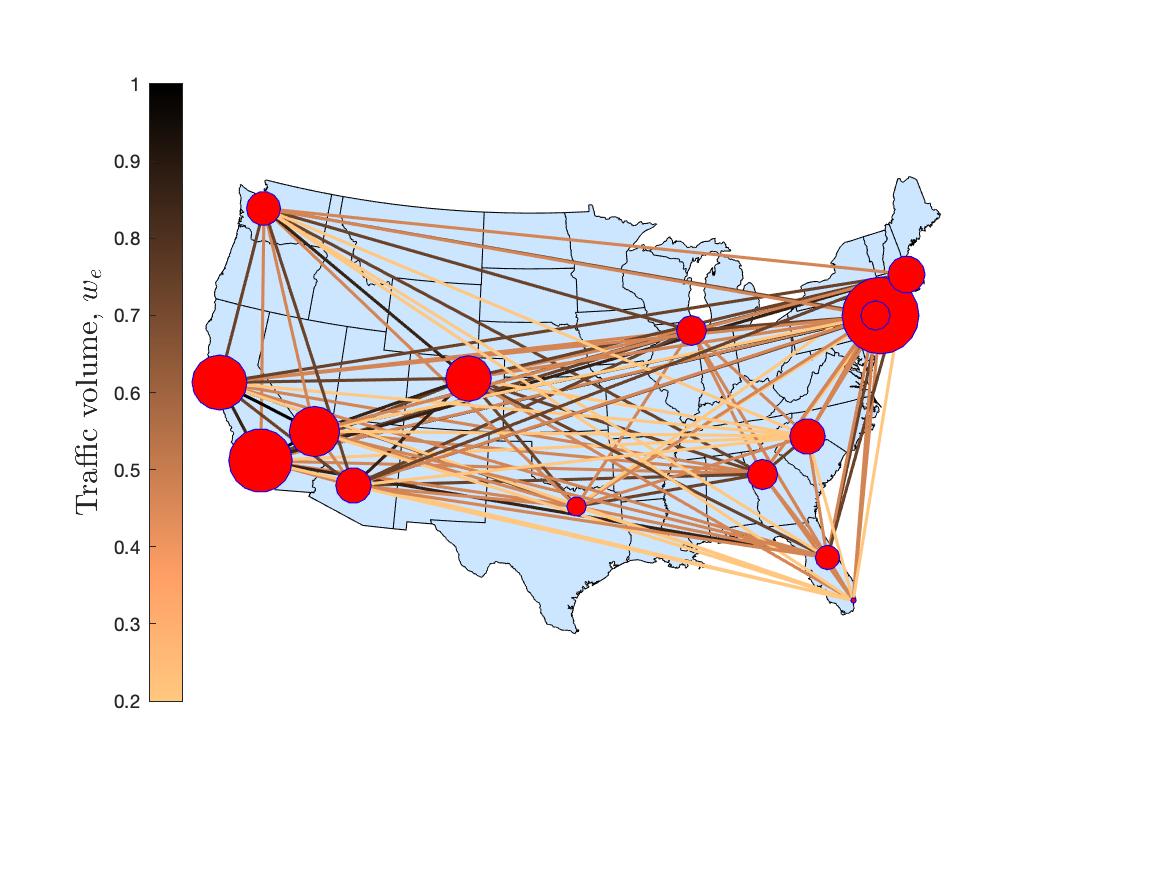}
    \subfigure[]{}
    \includegraphics[width=.28\linewidth,trim={5cm 8cm 8cm 0},clip]{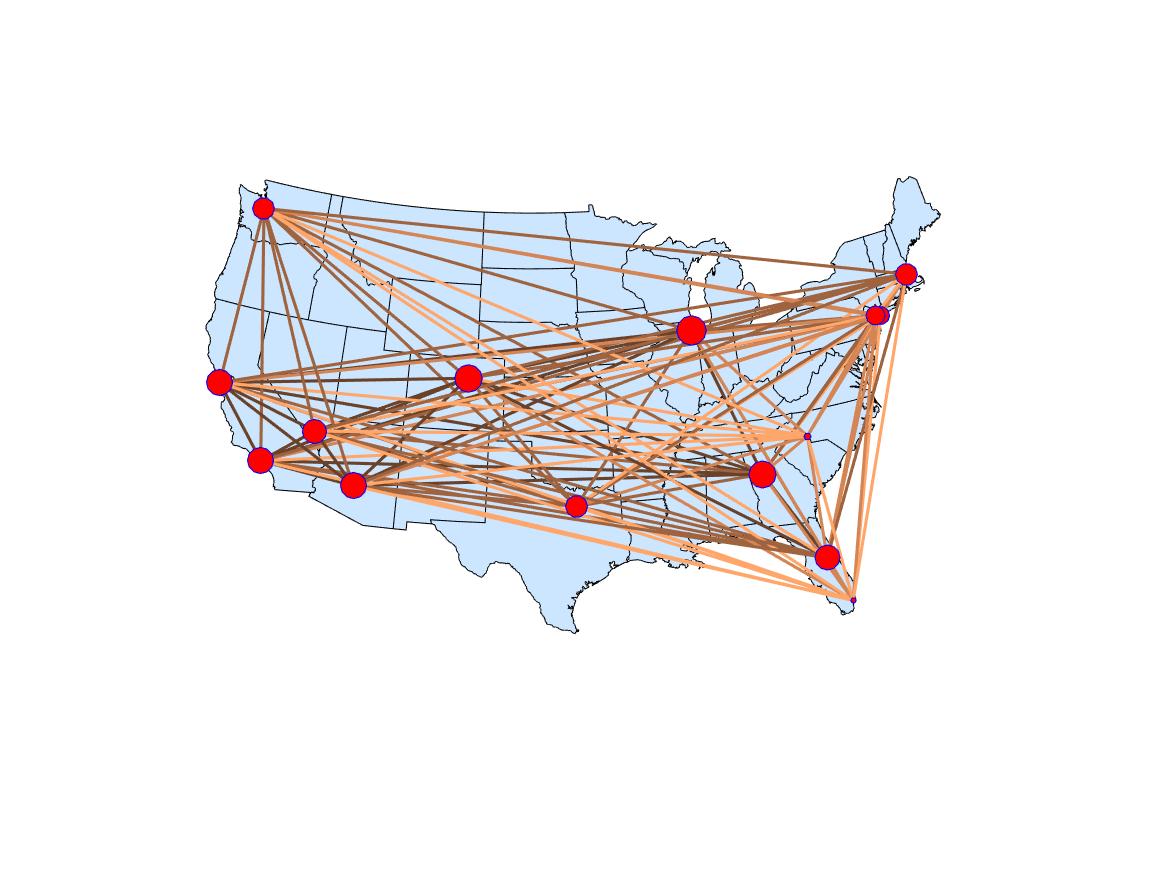}
    \subfigure[]{}
    \includegraphics[width=.28\linewidth,trim={5cm 8cm 8cm 0},clip]{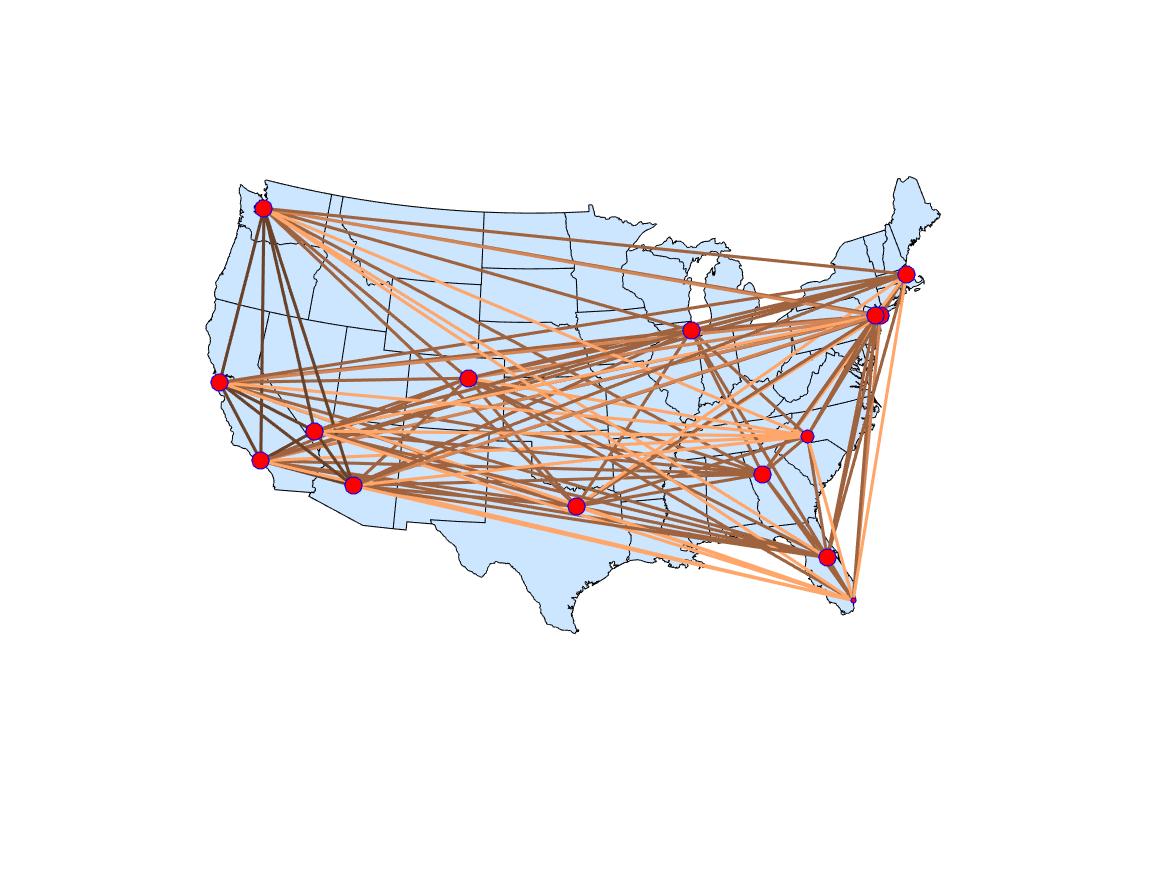}
\caption{(a) Meta-population network of 15 United States hubs and their normal traffic volume. All the sub-populations are experiencing $8$ days lag. The hubs are ranked based on their centrality index, $\eta_i$, which is reflected through the size of their indicating circles. The interconnections are ranked by their corresponding traffic volume which is specified by the color of edges.  (b) Optimal meta-population network of Fig.~\ref{Fig:Networks}(a) designed by the optimal approach \eqref{prob:opt1-1}, where a uniform noise distribution is applied. (c) Robust meta-population network of Fig.~\ref{Fig:Networks}(a) designed by the robust approach \eqref{prob:opt2-3}, where the worst case of applying the maximum noise input to the hub with highest centrality, New York,  is considered.}   \label{Fig:Networks} 
\end{figure*}

Air transportation plays an important role in introducing a new disease to a meta-population and spreading it within its sub-populations. Therefore, in this study a group of United States airports with the busiest airports is selected as the symbol of a real world epidemic network. The simulations are based on the dynamics \eqref{eq:system0} over the network of 15 hubs, see Fig.~\ref{Fig:Networks}, and their 104 weighted connections through air transportation. The United States air traffic data used in this study can be found in \cite{USdata:Online}.

Let us assume the a virus first arrives to the United States by one of these airports and infects this sub-population while spreading through other states by air transportation (Note that this scenario is just for the purpose of illustration and does not affect the generality of the studied epidemic problem). The resulting epidemic network is represented in Fig.~\ref{Fig:Networks}(a), where the hubs are experiencing $8$ days of delay and ranked based on the centrality index, $\eta_i$, of their airport.  
The scale of circles is correlated with centrality index of its corresponding airport. The interconnection between every pair of hubs is ranked based on traffic volume between their airports, which is specified by the color of links. Applying the proposed optimal traffic control method \eqref{prob:opt1-1} on this network will result in network Fig.~\ref{Fig:Networks}(b) with a lower range of centrality for all the hubs. Additionally, implementing the robust optimization approach \eqref{prob:opt2-3} on the original network will change the network structure to Fig.~\ref{Fig:Networks}(c). 

\begin{figure}
	\centering
	\includegraphics[trim={0 1cm 0 0}, width=0.75\textwidth]{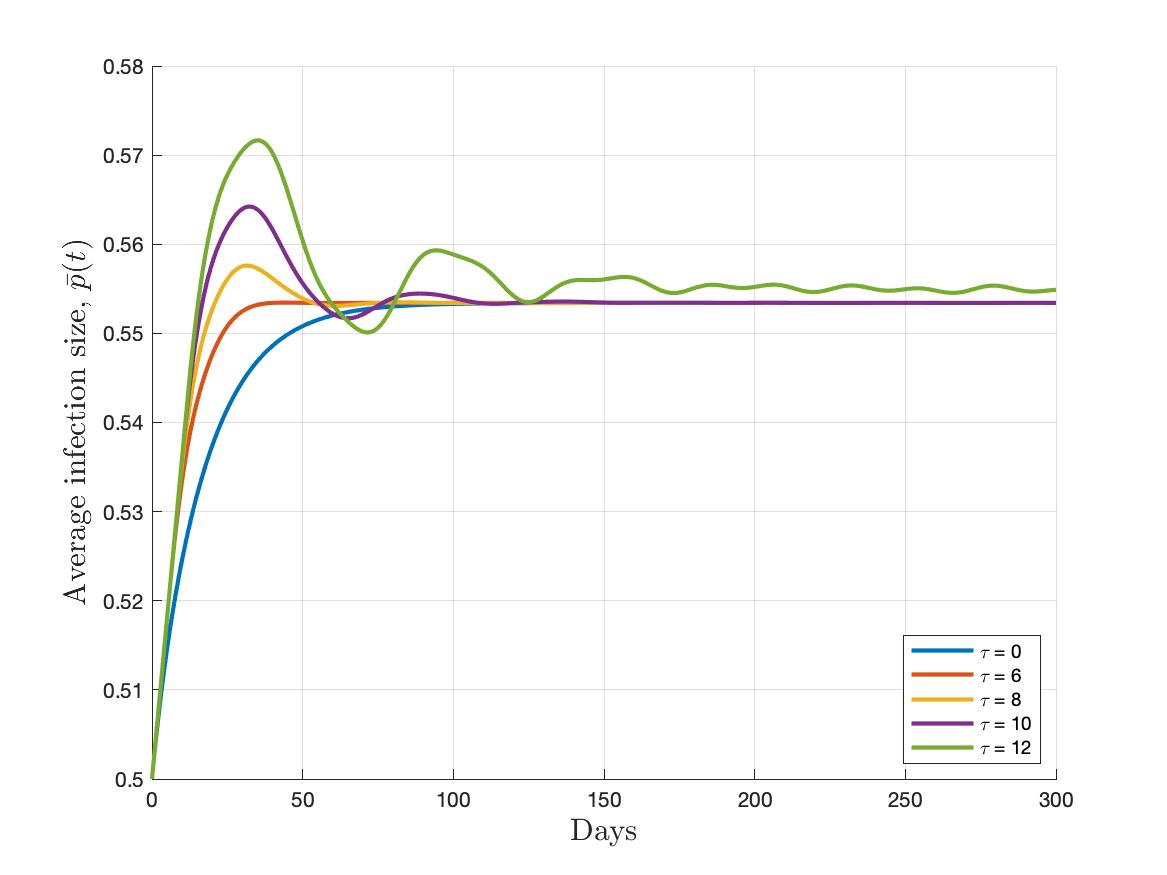}
	\caption{\small{The average infection size, $\bar{p}(t) = \frac{1}{n}\sum_{i \in \VV} p_i(t)$, of the meta-population network shown in Fig.~\ref{Fig:Networks}(a) with different time-delays. $50$ percent of the  meta-population is initially infected and $\mathcal{R}_0=2.3$. It is assumed that there is effective social distancing within every sub-population, $a_{ii}=0$ for $i=1,2,\ldots,n$..}}
	\label{Fig:Delayeffect}
\end{figure}

Fig.~\ref{Fig:Delayeffect} shows a comparison between the infection size of the simulated airport network when $50$ percent of the meta-population is initially infected and $\mathcal{R}_0=2.3$. For the network with no delay, $\tau=0$ days, or comparatively small delays, there would be no fluctuations in the epidemic dynamics but as the delay increases pulses begin to appear. 
Experiencing $12$ days of lag results in approximately $2$ percent (which corresponds to a considerable number of individuals in the United States population) increase in the average infection size within the first $30$ days of epidemic onset. 

\begin{figure}
	\centering
	\includegraphics[trim={0 1cm 0 0}, width=0.75\textwidth]{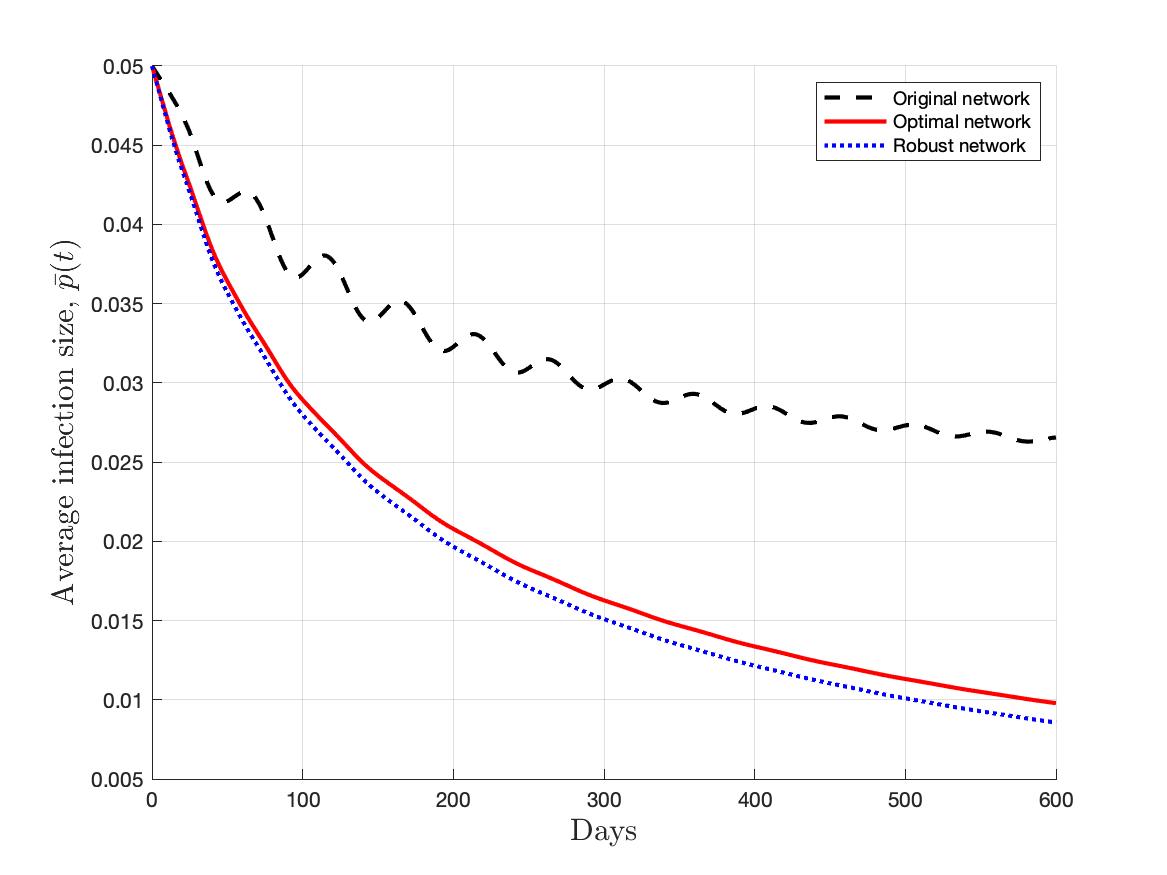}
	\caption{\small{Average infection size of the three networks in Figs.~\ref{Fig:Networks} with $\tau = 12$ days, $\mathcal{R}_0 \simeq 1.03$, and initial infection of $5$ percent.}}
	\label{Fig:Infection}
\end{figure}

The average infection size of the three networks Fig.~\ref{Fig:Networks} with $\tau=12$ days and initial infection of $5$ percent is presented in Fig.~\ref{Fig:Infection}, which indicates that the proposed methods are successfully decreasing the infection size. 

\begin{figure*}
    \centering
    \subfigure[]{}{
    \includegraphics[width=.31\linewidth,trim={2cm 1cm 2cm 0},clip]{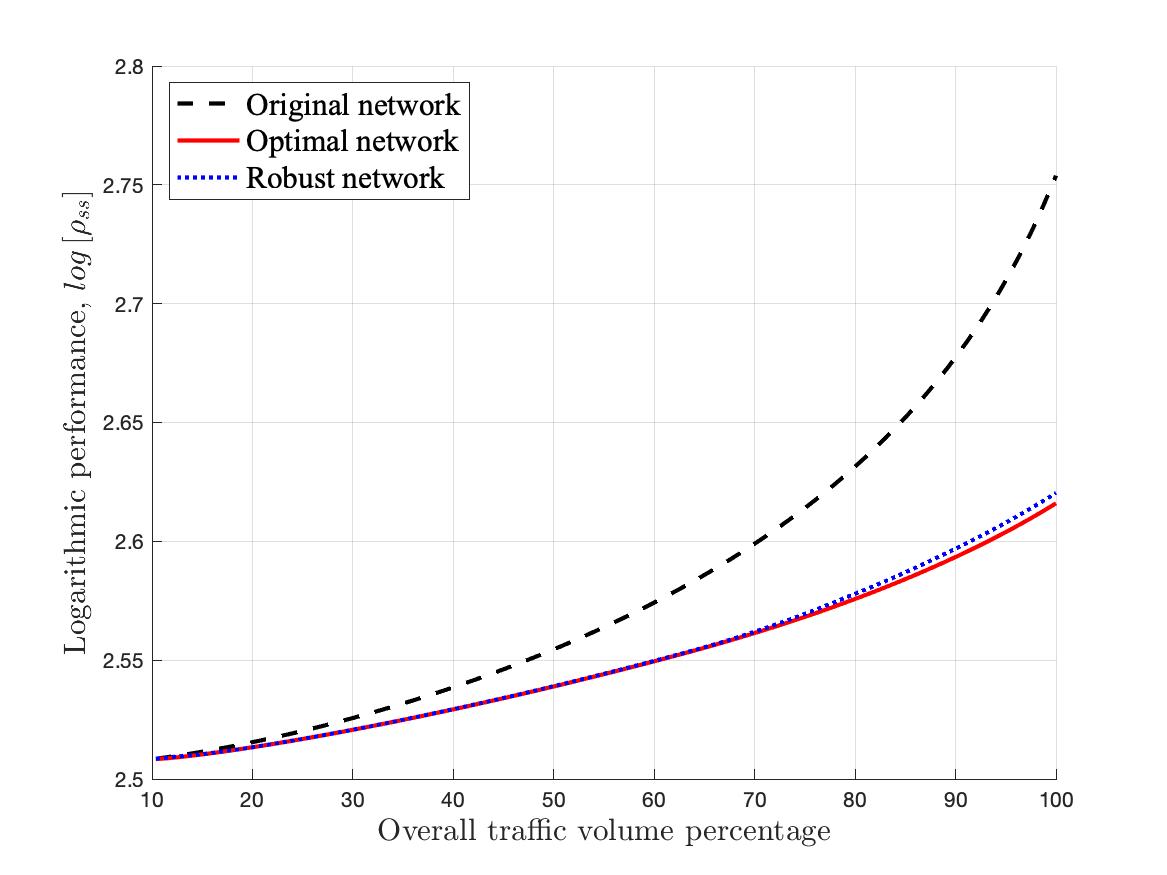}             
       }\hfill
    \subfigure[]{}{
    \includegraphics[width=.31\linewidth,trim={1cm 1cm 3cm 0},clip]{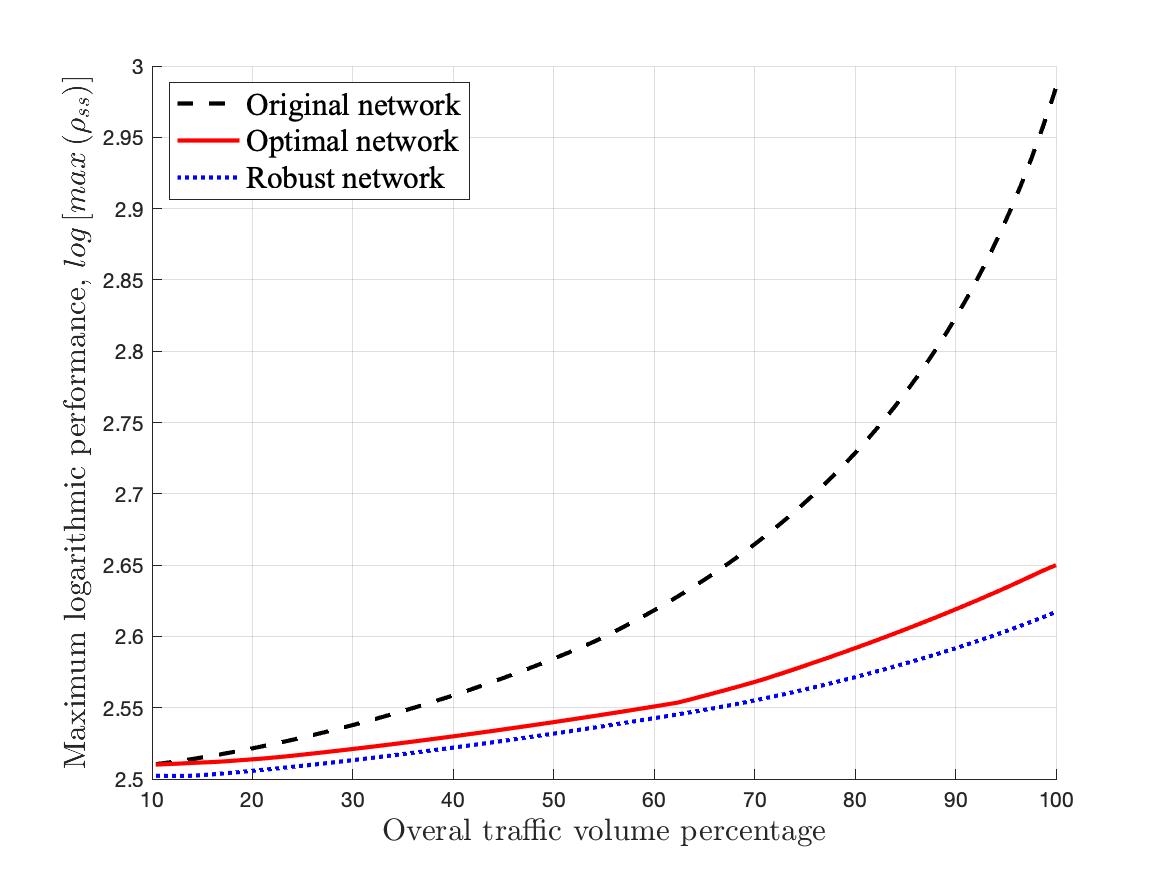}}\hfill
    \subfigure[]{}{
    \includegraphics[width=.31\linewidth,trim={0cm 1cm 3cm 0},clip]{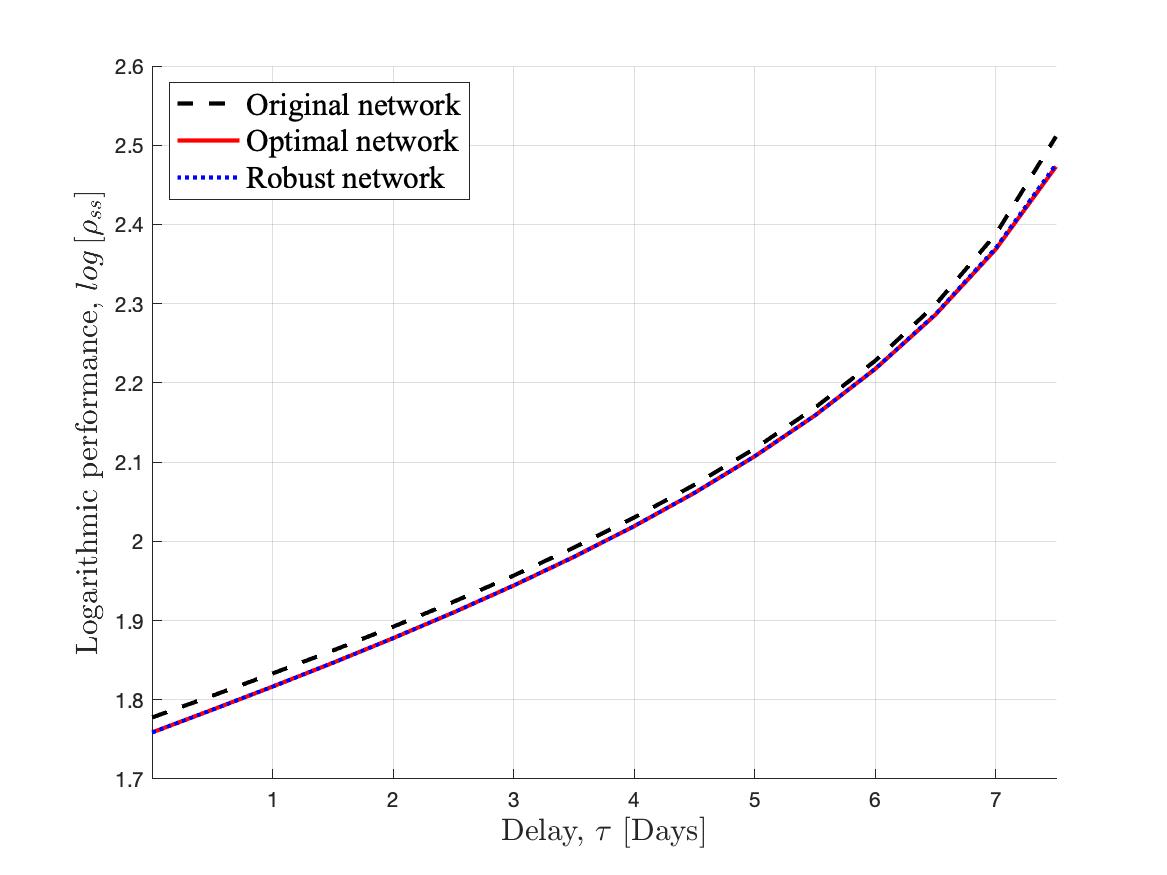}}
\caption{(a) Performance comparison between the original network Fig.~\ref{Fig:Networks}(a), the optimal network of Fig.~\ref{Fig:Networks}(b) and the robust network of Fig.~\ref{Fig:Networks}(c) in the case of uniform noise distribution. (b) Performance comparison between the original network Fig.~\ref{Fig:Networks}(a), the optimal network of Fig.~\ref{Fig:Networks}(b) and the robust network of Fig.~\ref{Fig:Networks}(c) in the case of extreme noise distribution. (c)  Logarithmic performance measure of networks Fig.~\ref{Fig:Networks} with respect to time-delay.}   \label{Fig:norm} 
\end{figure*}

The logarithmic scale of performance measure and its maximum for the networks in Fig.~\ref{Fig:Networks} are presented in Figs.~\ref{Fig:norm}(a) and \ref{Fig:norm}(b). In the case of uniform disturbance distribution, both optimal and robust controllers result in equal levels of performance for lower traffic percentages, while for high overall traffic volumes, over $70$ percent of normal volume, the optimal control performance decreases with a lower rate than that of robust control. However, in the worst case of noise distribution, robust controller shows a better performance, as it was expected. 

\begin{figure}
	\centering
	\includegraphics[trim={0 1cm 0 0}, width=0.75\textwidth]{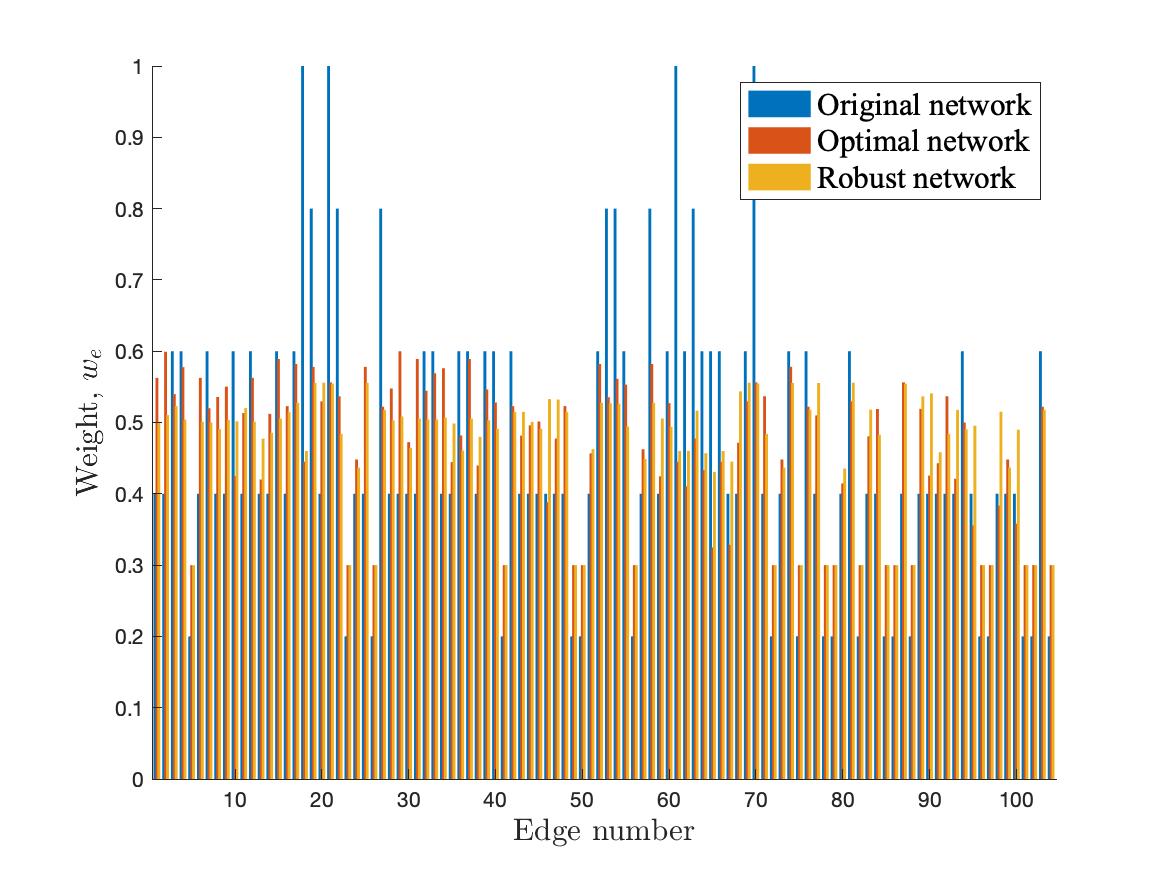}
	\caption{Weight (traffic volume) comparison between the original network Fig.~\ref{Fig:Networks}(a), optimal network Fig.~\ref{Fig:Networks}(b), and robust network Fig.~\ref{Fig:Networks}(c).}
	\label{Fig:Weightbar}
\end{figure}

\begin{figure}
	\centering
	\includegraphics[trim={0 1cm 0 0}, width=0.75\textwidth]{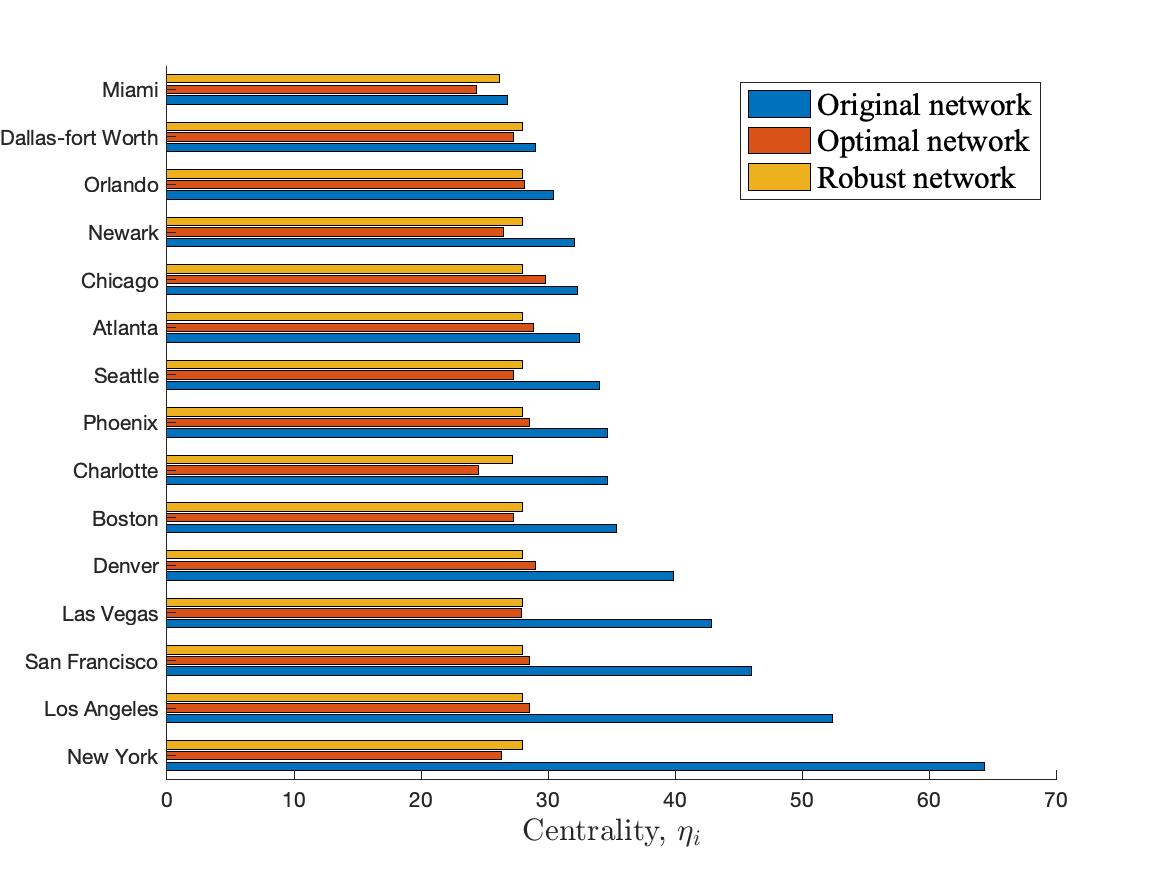}
	\caption{Centrality comparison between the hubs of original network Fig.~\ref{Fig:Networks}(a), optimal network Fig.~\ref{Fig:Networks}(b), and robust network Fig.~\ref{Fig:Networks}(c).}
	\label{Fig:Centralitybar}
\end{figure}

Figs.~\ref{Fig:Weightbar} and \ref{Fig:Centralitybar} present detailed information about the changes in traffic volume and centrality index of hubs, respectively. Unlike the core-periphery network, here the difference between optimal and robust network outputs is visible. The robust control tends to isolate more of the hubs with high centrality, which results in a better performance in worst cases, as shown in \ref{Fig:norm}(b). 

The detailed results of the performance improvement through optimal and robust epidemic controls are represented in Table.~\ref{table_2} for Case $1$ with uniform noise distribution, i.e. $\sigma_i=1$ for $i=1,\ldots,n$, and Case $2$ is the worst case explained in \eqref{worst-case}.
\begin{table}
\small
	\centering
	 \caption{The values and percentages of performance enhancement for the networks shown in Figs. \ref{Fig:Networks}(b) and \ref{Fig:Networks}(c) compared with Fig. \ref{Fig:Networks}(a)}
	\begin{tabular}{ |c|c|c| } 
\cline{2-3}
    \multicolumn{1}{c|}{} & Case 1 & Case 2\\ \hline 
     Original network (Fig.~\ref{Fig:Networks}(a))& 567 ~($0\%$)  &  966 ~($0\%$)\\ 
      \hline
     Optimal network (Fig.~\ref{Fig:Networks}(b))& 413 ~($+27\%$) & 446 ~($+54\%$)\\ 
      \hline
     Robust network (Fig.~\ref{Fig:Networks}(c))&  417 ~($+26\%$) & 414 ~($+57\%$)\\ 
     \hline
\end{tabular}
 \label{table_2}
\end{table}

\section{Discussions} \label{sec6}

In this study, the nonlinear and linear dynamics of an SIS network model in epidemic networks has been investigated. The studied meta-population is assumed to be experiencing delays due to the considerable proportion of pre-symptomatic population. The explicit centrality indices in the presence of model simplifications and testing errors have been derived and then used to develop optimal and robust traffic restriction methods for epidemic containment purposes. The proposed methods are implemented on a core-periphery network and a network of United States busiest airports. The simulation results indicate that the unavoidable delays in symptom development and infection identification can result in a significant difference in epidemic evolution for both cases, which requires more attention while designing potential government interventions. The proposed optimal and robust approaches, both based on the convex control method, shown to be capable of enhancing the delayed network's performance, and therefore, decreasing the infection rate considerably. Although adding more compartments to the network model increases the complexity of deriving explicit centrality indices, it can provide more realistic results. Implementing the proposed methods on the directed graphs with time-varying weights is also an interesting direction for improving the results of the current study.

Another aspect of an epidemic outbreak that requires more attention is \emph{vaccine tourism}. When vaccination becomes available for the first time, it motivates traveling to the centers offering it which results in the \emph{vaccine tourism} phenomenon. In the case of COVID-19 for instance, it is predicted that the first successful country in developing the vaccination will attract many tourists, at least until the vaccination resources are well distributed. With the vaccine distribution facing several challenges such as keeping vaccines at subzero temperatures, having low efficacy rate, producing limited dozes, etc, air transportation to the country offering the vaccine will probably be continued for several months. Therefore, most probably, with the \emph{vaccine tourism} comes a new wave of infection raise which requires a near-optimal and fast traffic volume control to mitigate its adverse consequences. The offered policies in this study can be generalized to meet the requirements for a safe and robust traffic control in the case of releasing a vaccination.


\bibliographystyle{ieeetr}
\bibliography{Bib}
 \end{document}